\documentclass[a4paper,11pt]{article}
\pdfoutput=1 

\usepackage{jcappub} 
\usepackage{xcolor}
\usepackage[T1]{fontenc} 
 
\usepackage{graphicx}
\usepackage{hyperref}
\usepackage{siunitx}
\usepackage{multirow}
\usepackage{txfonts}
\usepackage{subfigure}
\usepackage{float}
\usepackage{bm,array}

\title{\boldmath Kinetic Inductance Detectors \\ for the OLIMPO experiment: \\ in--flight operation and performance}

\author[a,b,1]{S. Masi,\note{Corresponding author.}}
\author[a,b]{P.~de~Bernardis,} 
\author[a,b]{A.~Paiella,}
\author[a,b]{F.~Piacentini,}
\author[a,b]{L.~Lamagna,}
\author[a,b]{A.~Coppolecchia,}
\author[c]{P.~A.~R. Ade,}
\author[a,b]{E.~S. Battistelli,}
\author[d]{M.~G.~Castellano,}
\author[d,e]{I.~Colantoni,}
\author[a,b]{F.~Columbro,}
\author[a,b]{G.~D'Alessandro,}
\author[a,b]{M.~De Petris,}
\author[f]{S.~Gordon,}
\author[g]{C.~Magneville,}
\author[f,h]{P.~Mauskopf,}
\author[d]{G.~Pettinari,}
\author[c]{G.~Pisano,}
\author[i]{G.~Polenta,}
\author[a,b]{G.~Presta,}
\author[i]{E.~Tommasi,}
\author[c]{C.~Tucker,}
\author[l,m]{V.~Vdovin,}
\author[i]{A.~Volpe}
\author[g]{and D.~Yvon}

\affiliation[a]{Dipartimento di Fisica, \emph{Sapienza} Universit\`a di Roma, P.le A. Moro 2, 00185 Roma, Italy}
\affiliation[b]{Istituto Nazionale di Fisica Nucleare, Sezione di Roma, P.le A. Moro 2, 00185 Roma, Italy}
\affiliation[c]{School of Physics and Astronomy, Cardiff University, Cardiff CF24 3YB, UK}
\affiliation[d]{Istituto di Fotonica e Nanotecnologie - CNR, Via Cineto Romano 42, 00156 Roma, Italy}
\affiliation[e]{{\sl current address}: CNR-Nanotech, Institute of Nanotechnology \\ c/o Dipartimento di Fisica, \emph{Sapienza} Universit\`a di Roma, P.le A. Moro 2, 00185, Roma, Italy}
\affiliation[f]{School of Earth and Space Exploration, Arizona State University, Tempe, AZ 85287, USA}
\affiliation[g]{IRFU, CEA, Universit\'e Paris-Saclay, F-91191 Gif sur Yvette, France}
\affiliation[h]{Department of Physics, Arizona State University, Tempe, AZ 85257, USA}
\affiliation[i]{Italian Space Agency, Roma, Italy}
\affiliation[l]{Institute of Applied Physics RAS, State Technical University, Nizhnij Novgorov, Russia}
\affiliation[m]{ASC Lebedev PI RAS, Moscow, Russia}

\emailAdd{silvia.masi@roma1.infn.it}

\abstract{We report on the performance of lumped--elements Kinetic Inductance Detector (KID) arrays for mm and sub--mm wavelengths, operated at \SI{0.3}{K} during the stratospheric flight of the OLIMPO payload, at an altitude of \SI{37.8}{km}. We find that the detectors can be tuned in--flight, and their performance is robust against radiative background changes due to varying telescope elevation. We also find that the noise equivalent power of the detectors in flight is significantly reduced with respect to the one measured in the laboratory, and close to photon-noise limited performance. The effect of primary cosmic rays crossing the detector is found to be consistent with the expected ionization energy loss with phonon--mediated energy transfer from the ionization sites to the resonators. In the OLIMPO detector arrays, at float, cosmic ray events affect less than $4\%$ of the detector samplings for all the pixels of all the arrays, and less than $1\%$ of the samplings for most of the pixels. These results are also representative of what one can expect from primary cosmic rays in a satellite mission with similar KIDs and instrument environment.}
\begin{document}
\maketitle
\flushbottom

\section{Introduction}

Kinetic Inductance Detectors (KIDs) potentially fulfill all the requirements for the next-generation Cosmic Microwave Background (CMB) space mission (see {\it e.g.} \citep{Delabrouille_CORE, deBernardis_CORE}): they are replicable in large--throughput arrays, are fabricated on solid silicon wafers and do not require, in their simplest implementation, membranes, nor other delicate microstructures. Thus, KIDs are relatively easy to fabricate, and mechanically robust. Moreover, they are fast, and recover in a short time from cosmic rays hits. Large KID arrays are now operated routinely at the IRAM \SI{30}{m} telescope (see {\it e.g.} \cite{2014AA...569A...9C, 2017AA...599A..34R, 2018AA...609A.115A}), and are going to be used in other ground--based mm--wave telescopes (see {\it e.g.} \cite{2016SPIE.9906E..3LO,2018JLTP..193..120A}). However, they have never been used in space. 

The effects of cosmic rays hits are the most important concern for the operation of CMB detectors in space (see {\it e.g.} \cite{1990NIMPA.294..328C, 2014AA...569A..88C, 2014JLTP..176..815M}), and extensive simulation/laboratory test activities have been carried out for KIDs in this respect (see {\it e.g.} \cite{2016AA...592A..26C, 2017AA...601A..89B, 2018AA...610A..45M, 2016SPIE.9914E..07G, backshort2}). However, a direct in--flight validation would be valuable, to test these detectors in the actual cosmic rays background. 

In this paper, we present for the first time an analysis of the actual performance of KIDs in a near--space environment: the stratospheric balloon altitude (\SI{37.8}{km}) achieved during the 2018 flight of the OLIMPO experiment (\citep{2005ESASP.590..581M, Coppolecchia2013}). This is timely since KIDs are going to be used in the near future in forthcoming balloon--borne experiments: in particular, larger arrays of KIDs covering shorter wavelengths than those covered by OLIMPO are ready to fly in the BLAST--TNG experiment \citep{2018SPIE10708E..0LL}. Moreover, KIDs start to be considered as a viable option for forthcoming CMB space missions, and even a near-space validation can be valuable when comparing different detection technologies.

A near--space validation of detectors to be used in a space mission is common practice. For example, the spider--web and polarization--sensitive detectors used in Planck--HFI (\citep{2010AA...520A...1T, 2011AA...536A...1P}) were tested in the BOOMERanG (\citep{2002ApJS..138..315P, 2003ApJS..148..527C, 2006AA...458..687M}) and Archeops \citep{2002APh....17..101B} stratospheric balloon flights. The balloon--borne measurements validated the spider--web and polarization--sensitive classes of bolometers, in terms of optical performance and sensitivity to cosmic rays hits.

The Lumped--Element KIDs of OLIMPO have been described in detail in \citep{Paiella2019}. Here we focus on the optimization carried out during the 2018 flight of OLIMPO, the measured in--flight performance, and the response to cosmic rays hits. We show that KIDs operate very well in near--space: no show--stoppers have been faced during extensive tests, confirming the potential of these detectors for the forthcoming near--space and space--borne surveys of the mm and sub--mm sky. 

\section{The KID arrays for OLIMPO}

OLIMPO is a balloon--borne telescope (\SI{2.6}{m} diameter aperture), feeding a differential Fourier transform spectrometer (DFTS) and four arrays of detectors, operating at 150, 250, 350, and \SI{460}{GHz}. The DFTS can be removed from the optical path by command, moving a relay mirror, resulting in an alternative 4-bands photometer configuration. The arrays fill the available focal plane area, which for a Ritchey--Chretien telescope like the OLIMPO one is quite small (covering $\sim 1^\circ$ of the sky, in diameter). For this reason, the number of optically independent pixels is limited. We have 19, 37, 23 and 41 active pixels, respectively for the 150, 250, 350, and \SI{460}{GHz} arrays. A few blank pixels (4 for the \SI{150}{GHz} array and 2 for each of the other arrays) have been added for monitoring purposes. Given the relatively high expected background on the OLIMPO detectors (see table \ref{tab:background}) the detectors have been optimized for operation at $\sim 0.3$K. Moreover, the arrays have been optimized to withstand variable background conditions. In fact, the radiative background of OLIMPO increases by a factor $\sim 2$ to $3$ when switching from the photometer configuration, when the arrays look directly to the telescope, to the spectrometer configuration, when the arrays look at the telescope through the mirrors and the wire--grids of the ambient--temperature differential spectrometer. 

Details of the design optimization of KIDs are reported in \citep{Paiella2017_ISEC, Paiella2019}. The absorbers geometry is a IV order Hilbert curve, where the base length scales with the observed frequency. The detectors are front illuminated via single--mode waveguides for the 150 and \SI{250}{GHz} arrays and via single--mode flared waveguides for the 350 and \SI{460}{GHz} arrays. The detector arrays have been made with \SI{30}{nm} thick aluminum (with $T_{c}=\SI{1.31}{K}$ \citep{Paiella2016}) deposited on silicon substrates, with diameter depending on the diffraction-limited field of view, and thickness depending on the observed wavelength: 3''  diameter $\times\SI{135}{\mu m}$ thickness for the \SI{150}{GHz} array, 3''$\times \SI{100}{\mu m}$ for the \SI{250}{GHz} array, 2''$\times\SI{310}{\mu m}$ for the \SI{350}{GHz} array, and 2''$\times\SI{135}{\mu m}$ for the \SI{460}{GHz} array. In order to act as a backshort and a phonon trap, the face of the silicon substrate opposite to the detectors has been metalized with \SI{200}{nm} of aluminum. 

Two identical readout electronics are used, biasing and reading the 150 and \SI{460}{GHz} arrays and the 250 and \SI{350}{GHz} arrays respectively. Each readout produces a comb of frequencies matching the resonant frequencies of the pixels of the two arrays being biased \cite{2016JAI.....541003G}. The \emph{cold} and the \emph{room--temperature} electronics have been optimized at ground in terms of noise and power budget \citep{Paiella2018_WOLTE13}. The \emph{client} software running on the on--board computer of OLIMPO allows to feed each resonator with tailored bias power and frequency. Detectors can be tuned during the flight via commanding, and monitored via the telemetry downlink. We prepared two sets of bias power for each readout line, one for the photometer configuration and one for the spectrometer configuration, both optimized at ground with an optical load similar to the one expected for the stratospheric environment. The flight software is able to switch between the two sets when the experiment switches between the two configurations.

\section{Test conditions}

The OLIMPO payload was launched at 07:07 GMT on July 14$^{\rm th}$, 2018, and reached the float altitude (\SI{37.8}{km}) at 12:30 GMT of the same day. All the measurements described in this paper were carried out in the first 24 hours of the OLIMPO flight, when the fast line-of-sight telemetry was available.

The detector readout system operated flawlessly during the flight. The power dissipation scheme developed for the power components in the readout boards, based on copper straps thermally connecting the power--hungry chips to a large radiator, maintained the temperature of the chips in their operating range, despite of the lack of air cooling. The temperature of the FPGAs at float were constantly around $\SI{60}{^{\circ}C}$.

The performance of KIDs is strongly affected by the temperature of the array wafers, and by the radiative background. Both can change during a stratospheric balloon flight, due to launch operations, reduction of atmospheric emission with altitude, LN$_{2}$ and L$^4$He temperature drifts with increasing altitude, instrument configuration changes (photometer vs. spectrometer), and change of the residual atmospheric background with telescope elevation. Moreover, radiative background variations can induce changes in the temperature of the detector array wafers, so that the detector performance is affected in a mixed way. 

In the following we describe the relevant changes of temperature and background conditions during the OLIMPO flight, their monitoring and their effects.

\subsection{Radiative background on the detectors} \label{radiative}

\newcolumntype{C}{>{\centering\arraybackslash}p{5em}}
\begin{table}[htb]
	\centering
		\fontsize{10pt}{18pt}\selectfont{
		\begin{tabular}{l|c|c|c|c}
		\hline
		\hline
		\multicolumn{1}{c|}{\multirow{1}{*}}&
		\multicolumn{4}{c}{\multirow{1}{*}{ Background $\left[\SI{}{pW}\right]$ }}\\
		\cline{2-5}
		\multicolumn{1}{l|}{\multirow{1}{*}{Source}}&
		\multicolumn{1}{c|}{\multirow{1}{*}{ \SI{150}{GHz} }}&
		\multicolumn{1}{c|}{\multirow{1}{*}{ \SI{250}{GHz} }}&
		\multicolumn{1}{c|}{\multirow{1}{*}{ \SI{350}{GHz} }}&
		\multicolumn{1}{c}{\multirow{1}{*}{ \SI{460}{GHz} }}
		\\
		\hline
		\hline
CMB & 0.41 & 0.57 & 0.03 & 0.02 \\
atmosphere ($e=45^\circ$) & 0.04 & 0.3 & 2.5 & 13 \\
atmosphere ($e=15^\circ$) & 0.11 & 0.8 & 7.0 & 36 \\
primary mirror & 0.8 & 4.7 & 1.6  & 3.0 \\
secondary mirror & 0.8 & 4.7 & 1.6 & 3.0 \\
spectrometer & 9.5 & 60 & 20 & 40 \\
window & 1.7 & 8.4 & 2.3 & 3.9 \\
\SI{60}{K} filters & 0.5 & 2.6 & 0.7 & 1.1 \\
\SI{30}{K} filters & 0.2 & 0.9 & 0.2 & 0.4 \\
\SI{2}{K} filters & 0.0 & 0.0 & 0.0 & 0.0 \\
\SI{0.3}{K} filters & 0.0 & 0.0 & 0.0 & 0.0 \\
\SI{0.3}{K} environment & 0.2 & 0.07 & 0.03 & 0.02  \\
\hline
TOTAL Photometer ($e=45^\circ$)  & 2.7 & 14 & 5 & 13  \\
TOTAL Spectrometer ($e=45^\circ$)  & 8 & 50 & 15 & 30  \\
\hline
TOTAL Photometer ($e=15^\circ$)  & 2.7 & 14 & 7 & 25  \\
TOTAL Spectrometer ($e=15^\circ$)  & 8 & 50 & 17 & 43  \\
\hline
		\hline
		\multicolumn{1}{c|}{\multirow{1}{*}}&
		\multicolumn{4}{c}{\multirow{1}{*}{ Optical photon--noise limit $\left[\SI{}{\mu K \sqrt{\rm s}}\right]$ }}\\
		\hline
NET$_{\rm RJ}$ -- Photometer ($e=45^\circ$)   & 65 & 30 & 65 & 65 \\		
NET$_{\rm CMB}$ -- Photometer ($e=45^\circ$) & 115 & 325 & 780 & 2900 \\		
		\hline
NET$_{\rm RJ}$ -- Spectrometer ($e=45^\circ$)   & 115 & 55 & 120 & 104 \\		
NET$_{\rm CMB}$ -- Spectrometer ($e=45^\circ$) & 190 & 550 & 1400 & 5100 \\		
		\hline
		\hline

		\end{tabular}		
		}
		\caption{\small Radiative background and photon-noise-limits expected on the OLIMPO detectors for the photometer and spectrometer configurations. The total background is computed summing the contribution from each source times the product of the transmissions of the filters encountered by the radiation from that source before reaching the detectors. We assume the detector absorption efficiency measured in \cite{Paiella2019}. }
	\phantomsection\label{tab:background}
\end{table}

The radiative background on the detectors in a given spectral band depends on the pointed target, the emission of the residual atmosphere (which in turn depends on the payload altitude and the telescope elevation angle $e$), the emission of the optical elements (telescope, and spectrometer if inserted) and of the window of the cryostat (made of high density polyethylene \citep{DALESSANDRO201859}) and the emission of the cold optical elements (filters and mirrors) inside the cryostat. In table \ref{tab:background} we report our best estimates of the photon background and noise during the flight, for the two different configurations (photometer and spectrometer) of OLIMPO. From this table it is evident that the largest background changes expected during the OLIMPO mission are due to the insertion of the differential Fourier transform spectrometer. Atmospheric background changes are expected to be significant mainly in the two higher frequency bands. 

\subsection{$^3$He fridge temperature}

The OLIMPO KIDs arrays are thermally connected to a $^3$He evaporation refrigerator, pre--cooled by a \SI{60}{l} superfluid $^4$He bath, surrounded by a shield cooled by a \SI{74}{l} liquid nitrogen tank. The $^4$He tank is pumped on the ground by means of a rotary pump. It is then sealed just before the balloon launch, using a motorized valve, and remains sealed during ascent. At float the motorized valve is opened and the $^4$He bath is connected to the low pressure of the stratosphere ($\sim$ \SI{4}{mbar} at the float altitude). The motorized valve was closed by command at 05:57 GMT of July 14th, in preparation of the launch, and was reopened by command at 11:37 GMT of July 14th, at an altitude of $\sim$ \SI{35}{km}. During ascent the vapor pressure in the sealed $^4$He tank increased, and, as a consequence, the $^4$He bath temperature increased, together with the temperature of the $^3$He evaporator: from \SI{1.71}{K} and \SI{307}{mK} at launch to \SI{1.90}{K} and \SI{320}{mK} just before re--opening the motorized valve. At float, when the motorized valve was opened, the pressure started to decrease, and both the $^4$He bath and the $^3$He evaporator temperatures drifted down, getting to \SI{1.65}{K} and \SI{295}{mK} in a few hours. In the subsequent operation, the temperature of the $^3$He evaporator remained stable at \SI{295}{mK} during photometric observations, and responded to the increased background during spectroscopic observations, rising by $\sim \SI{0.5}{mK}$ to \SI{1}{mK}. Within each $\sim$ 1 hour long observation batch, the temperature of the refrigerator remained stable within \SI{0.1}{mK}. 

The temperature of the $^3$He evaporator depends slightly on the elevation pointing of the telescope: the cryostat is vertical when the elevation is set to $e=0^\circ$, and is increasingly tilted as the elevation is increased. At launch, the telescope boresight elevation was set at $e=3^{\circ}$ (``safe'' position). It was moved to $e=15^{\circ}$ when the payload reached an altitude of $\sim$ \SI{32}{km}. 

\subsection{Detector arrays temperature variations}

At float, the $^3$He evaporator stabilizes at around \SI{295}{mK}, and the arrays stabilize at a slighty higher temperature due to the thermal resistance between the arrays and the refrigerator. The temperature difference, as measured in the laboratory, is a few mK for the 150 and \SI{250}{GHz} arrays, $\sim$ \SI{30}{mK} for the 350 and \SI{460}{GHz} arrays.

When we switch from photometric to spectroscopic measurements, the in--band radiative background increases up to a factor of 3 (see section \ref{radiative}), but out--of--band contributions are also expected and do increase with the switch, thus contributing to the $^3$He evaporator temperature increase. As a consequence, the temperatures of the arrays and of the $^3$He fridge rise. The former rises by a larger amount than the latter due to the series of thermal resistances between the arrays wafers and the $^3$He refrigerator. We do not have thermometers on the arrays, in--flight, but we can analyze \emph{blank} (not illuminated) pixels, the response of which can be used to estimate the temperature variation of the detector arrays. 

\begin{figure}[h]
\centering
\includegraphics[scale=0.5]{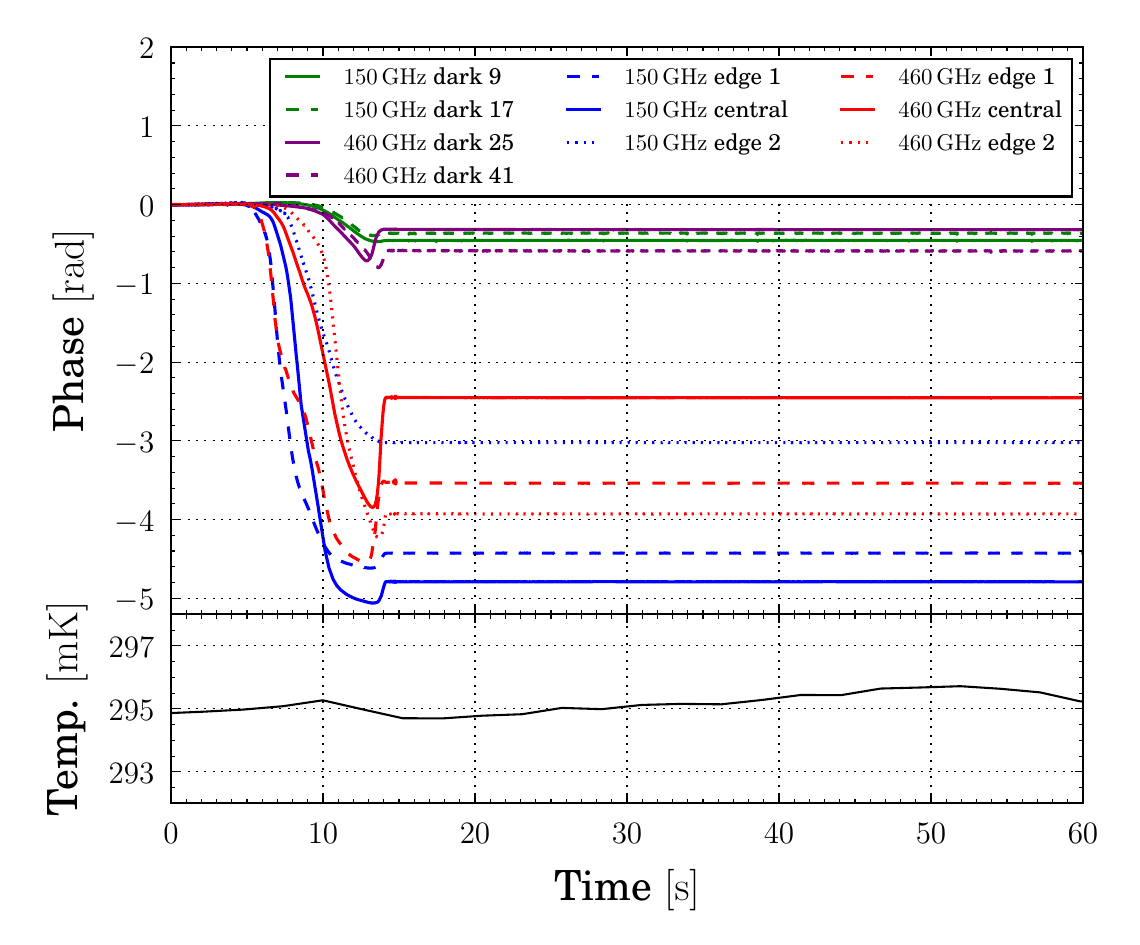}
\caption{\small \emph{Top panel}: \SI{60}{s} long KIDs phase-signal timestreams, acquired during the in--flight switch from photometric to spectroscopic configuration. The insertion of the relay mirror starts at $t=\SI{6}{s}$ and is completed at $t=\SI{15}{s}$. Green and purple lines are the phase signals for the {\it dark} pixels of the 150 and \SI{460}{GHz} array, respectively. Dashed, solid and dotted blue (red) lines are for the first, central and last pixel illuminated by the background change of the \SI{150}{GHz} (\SI{460}{GHz}) array, respectively. \emph{Bottom panel}: $^3$He evaporator temperature record during the same event. The sampling interval for the temperatures ($\sim$ \SI{2}{s}) was much slower than the sampling of the detector signals ($\sim$ \SI{8}{ms}).}
\phantomsection\label{fig:fotom_vs_spectro}
\end{figure}

In Figure \ref{fig:fotom_vs_spectro} we plot selected signals from \emph{active} and \emph{dark} pixels during the transition from photometric to spectroscopic measurements. These are compared to a simultaneous record of the $^3$He evaporator  temperature. We have plotted the timestreams for the {\it dark} pixels of the 150 and \SI{460}{GHz} arrays, and those at the {\it edges} and the {\it centers} of the same arrays. The switch from photometer to spectrometer is obtained moving a relay mirror in the optical path, to redirect the incoming radiation into the spectrometer. This operation requires $\sim$ \SI{10}{s} to complete (from $t=\SI{6}{s}$ to $t=\SI{15}{s}$ in  figure \ref{fig:fotom_vs_spectro}), with different pixels of the array sequentially illuminated by the increased background during the motion of the mirror. As {\it edge} pixels, we selected the first and the last pixels illuminated by the background change during the insertion of the relay mirror. From figure \ref{fig:fotom_vs_spectro} it is evident that the variation of the signals measured on the {\it dark} pixels happens after the complete insertion of the DFTS, and at the same time for all the {\it dark} pixels, even if they are diametrically opposite in the array. Therefore, these signals can not be interpreted as an optical or electrical cross--talk: they are solely due to a temperature variation of the array. Such a temperature variation can be estimated computing the ratio between the measured signals and the temperature responsivity of the {\it dark} pixels (see below): they are about \SI{7}{mK} for the \SI{150}{GHz} array and \SI{12}{mK} for the \SI{460}{GHz} one. These temperature variations are smaller than those measured in the laboratory, due to the reduced radiative background in flight. 

\section{In--flight KIDs performance} \label{optimization}

\subsection{In--flight tuning of the detectors} \label{tuning}

Detectors tuning consists of finding the resonant frequency and setting the bias frequency and power level for each detector, so that its performance (NET) is optimized in the actual operating conditions. 

During the part of the flight of interest here, communications with the experiment were provided by the line--of--sight telemetry (\SI{500}{kbps}, bidirectional), which allowed us to control in real--time the optimization of the working points of the detectors. So, we were able to tune the detectors, changing the frequency and power level of the bias excitation for each pixel, basically as in the laboratory. The tuning procedure started by producing the pre--programmed frequency comb with some nominal power levels; a \SI{200}{kHz} wide frequency sweep around these frequencies allowed us to find the resonances for all pixels; these values were finally set as the new comb of frequencies. At this point, we checked all the $S_{12}$ parameters near resonance, looking for deviations from the expected circular shapes. A few of them were found, and fixed by reducing the power level. The process took about 1 hour to tune the 120 working pixels of the 4 arrays. The optimization was simplified by the particular tradeoff between dynamic range and responsivity selected for the OLIMPO arrays (see \cite{Paiella2019} for details). 

\begin{figure}[!h]
\centering
\includegraphics[scale=0.39]{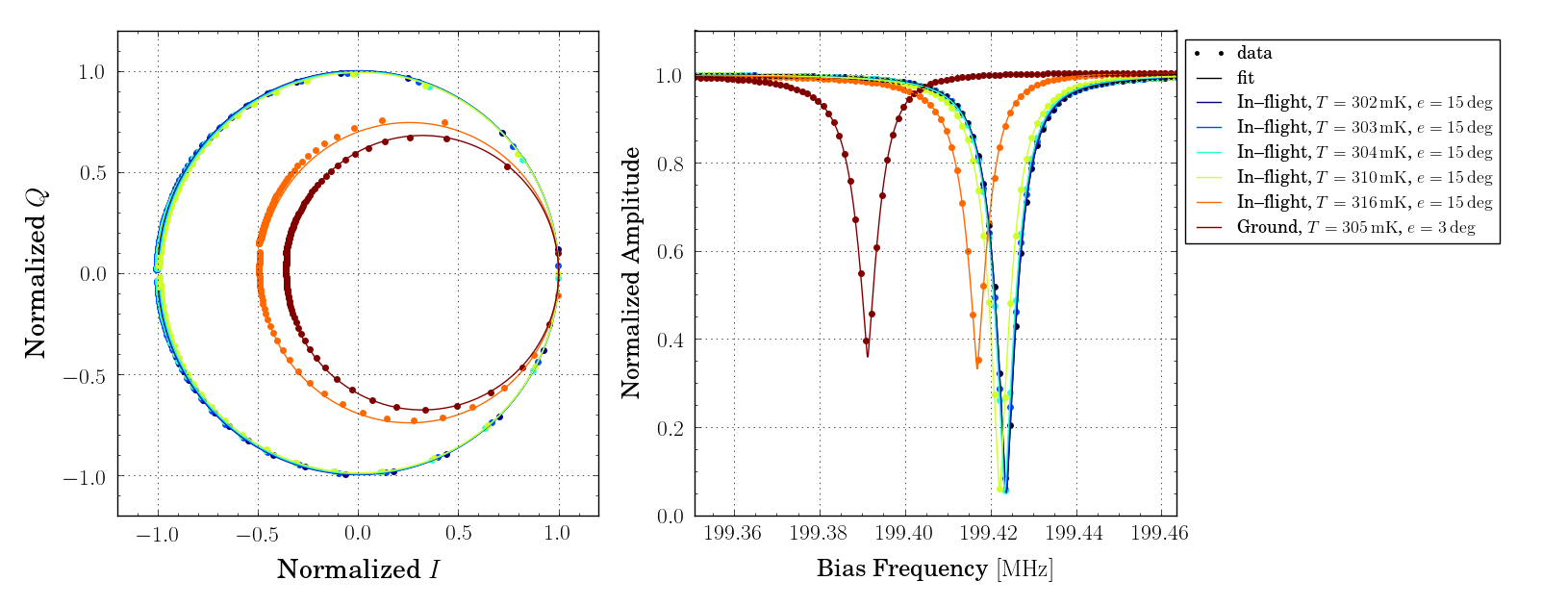}
\vspace{-0.8cm}
\caption{\small Resonator $S_{12}$ parameter in the complex plane (\emph{left}) and transmitted amplitude (\emph{right}) for pixel \#7 of the \SI{150}{GHz} array, for different fridge temperatures and background powers. The dots represent the actual measurements while the solid lines are the best fits. Flight and laboratory measurements are compared as detailed in the box.}
\phantomsection\label{fig:res7}
\end{figure}

In figure \ref{fig:res7} we compare, as a representative example, the tuning of pixel \#7 of the \SI{150}{GHz} array, at ground and in--flight. We report the resonance circles and the transfer function amplitudes for different $^3$He evaporator temperatures and also for different background power levels, plotting measurements taken in the laboratory and in the stratosphere, pointing at an elevation of 15$^\circ$. 

 Under similar background conditions, the resonant frequency of the KID resonator moves to lower frequencies when increasing the working temperature, as expected due to the increase of the kinetic inductance. During the flight, we measured the resonator response at different fridge temperatures and at an elevation of 15$^\circ$. The main difference between the ground and the in--flight resonances is due to the background power on the detectors, which decreases considerably because of the decrease of the temperatures of the window (the air temperature decreases by increasing the altitude) and the filters cooled by LN$_{2}$, and mainly because of the very low optical depth of the atmosphere at the altitude of the experiment, for all telescope boresights. Therefore the flight resonances have higher resonant frequencies, independent of their temperature. This plot already indicates a significant improvement of the performance of the detectors in flight, which is quantified below.

\subsection{In--flight performance verification} 

In order to control the behaviour of the arrays at different telescope boresight elevations, we checked both small and large changes in elevation. 

We find that the background changes produced in all the active pixels of the arrays by a $\sim$ 1$^\circ$ change in the elevation ({\it e.g.} from $e = 15^\circ$ to $ e = 16^\circ$), produce phase signal changes of the order of \SI{0.05}{rad}. Such relatively small changes do not require any retuning of the detectors. Larger elevation changes ({\it e.g.} from $e =16^\circ$ to $e = 5^\circ$) produce, instead, larger phase signal changes, of the order of \SI{1}{rad}, which do modify the responsivity of the detectors. However, due to the tradeoff between dynamic range and responsivity chosen during the design optimization of the OLIMPO detectors, there is no real need of retuning, if the detector responsivities are monitored using the signals from the calibration lamp (see section \ref{callamp}).

Larger background changes are induced by the insertion of the differential Fourier transform spectrometer (see table \ref{tab:background}). Here retuning is needed. For this reason, two different tuning tables (including bias frequencies and amplitudes for all pixels) have been optimized for the photometric and spectroscopic measurement modes, and are switched when switching from one observation mode to the other.

\subsection{Calibration Lamp signals} \label{callamp}

\begin{figure}[htb]
\centering
\includegraphics[scale=0.68]{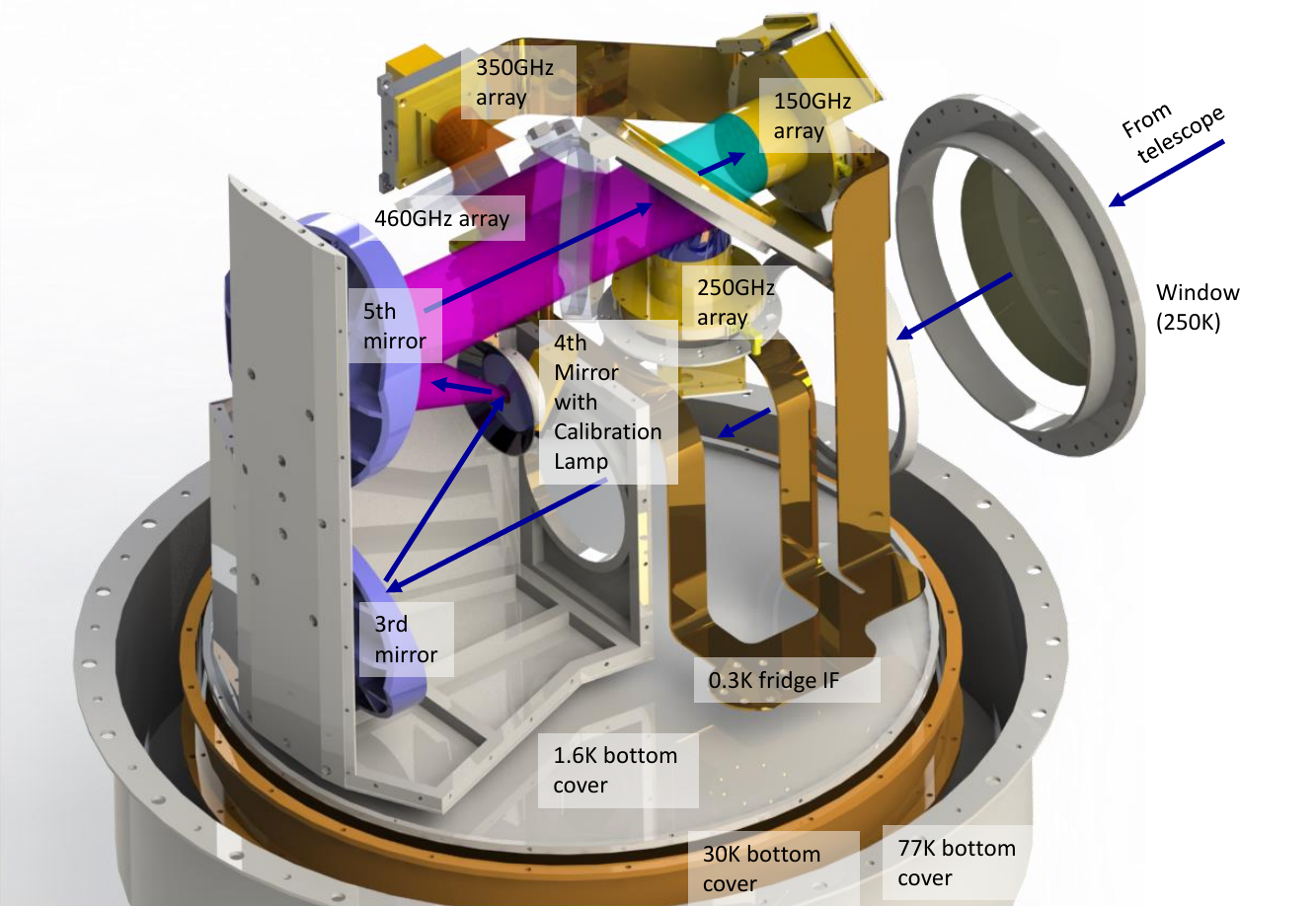}
\caption{\small Sketch of the OLIMPO cryogenic reimaging optics (3$^{\rm rd}$ mirror, 4$^{\rm th}$ mirror, 5$^{\rm th}$ mirror, dichroics and detector arrays). The cryogenic liquid tanks and many parts of the outer shell and shields have been hidden to show the optical system. The calibration lamp is located in the center of the 4$^{\rm th}$ mirror, which is the Lyot stop of the optical system, and close to the foci of the 3$^{rd}$ and 5$^{\rm th}$ mirrors. The beam from the calibration lamp, illuminating the four detector arrays is also shown. The path of the chief ray coming from the telescope and reaching the \SI{150}{GHz} array is indicated by the blue arrows. For scale, the diameter of the \SI{1.6}{K} L$^{4}$He tank is \SI{45}{cm}.}
\phantomsection\label{fig:callamp}
\end{figure}

A calibration transfer lamp, which can be activated by command or by timer, is placed in the center of the Lyot stop of the OLIMPO reimaging optics at \SI{1.6}{K} (see figure \ref{fig:callamp}).

The emitter is described in detail in  \cite{2007NIMPA.575..412A}, and is similar to a composite bolometer. When activated, driving a current through a resistor, it heats up an absorbing surface from \SI{1.6}{K} to a few tens of Kelvin. The emitter is inserted in an integrating cavity, feeding a Winston cone (\SI{5}{mm} aperture at the source side, \SI{12.7}{mm} aperture at the emission side, \SI{20.6}{mm} length). The emission--side aperture of the cone is located at the center of the Lyot stop (4$^{\rm th}$ mirror of the optical chain) of the OLIMPO Offner reimaging system, near the focus of the $5^{\rm th}$ mirror, thus illuminating with an approximately parallel beam all the detectors of the four focal plane arrays, with a reproducible thermal signal. The FWHM aperture of the beam emitted by the Winston cone is $\sim 40^\circ$, and the approximate top--hat response is flat to better than 10\% of the maximum for off axis angles $<14^\circ$, nominally corresponding to a roughly flat illumination beam area, wider than the array area. This is enough to warranty a good signal to noise ratio of the calibration lamp signals for all pixels of all arrays, so that responsivity changes can be monitored accurately for each pixel. However, as confirmed by GRASP simulations, the illumination of the focal plane shows a variation of a factor $\sim 2$ across the focal plane area. This is not flat enough to cross-calibrate different pixels assuming the same illumination from calibration lamp pulses for all pixels. 

During pre--flight laboratory calibrations (see \citep{Paiella2019}), the optical responsivity of all the pixels ${\cal R}_{lab,i}$ , for $i=1\dots N_{pix}$, is estimated measuring the signal produced by a chopped thermal source (\SI{300}{K} blackbody versus \SI{77}{K} blackbody, filling the beam, and seen through a neutral density filter with $\sim 1\%$ transmission and a thick--grill plate with $\sim 5\%$ transmission). In the same conditions, the response of all pixels to the calibration lamp $S_{lab,i}$ is also recorded, so that we can use it as a calibration transfer. Measuring the calibration lamp signals for all pixels in flight, $S_{flight,i}$, we can estimate the in--flight optical responsivity as 
$$
{\cal R}_{flight,i}={\cal R}_{lab,i} \frac{S_{flight,i}}{S_{lab,i}}
$$

This estimate of the responsivity of the detection system is useful to check if the in--flight performance has changed with respect to the laboratory one, and also to compare to the one obtained with astrophysical sources, in order to estimate the coupling efficiency between the telescope and the detection system. 

\begin{figure}[htb]
\centering
\includegraphics[scale=0.72]{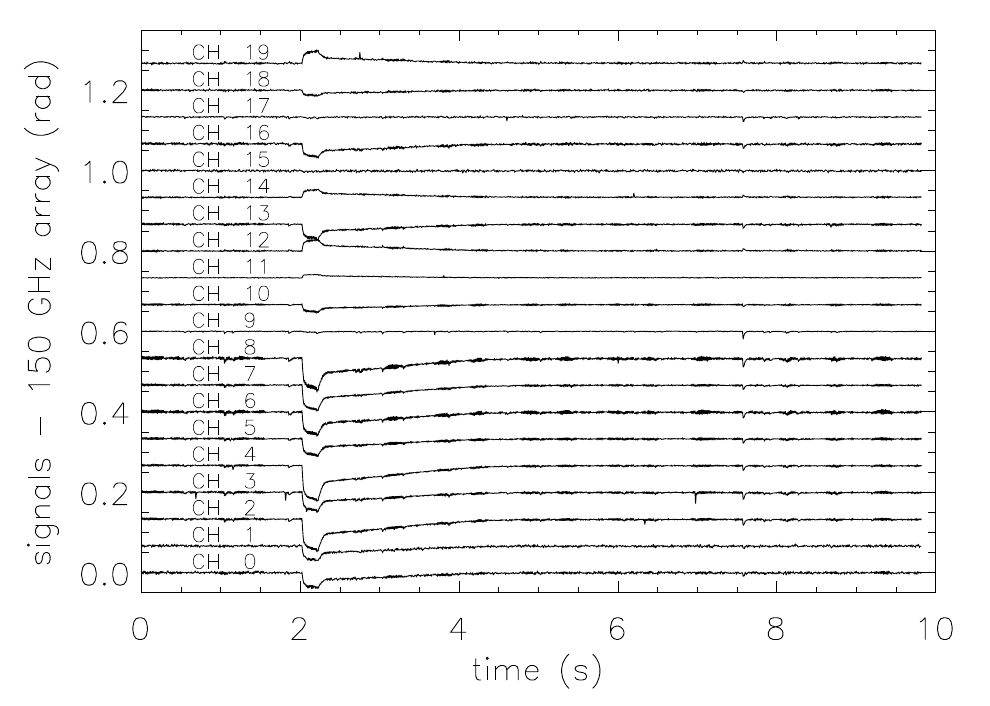}
\caption{\small A sample section of the raw signal tracks from 20 pixels of the \SI{150}{GHz} array, arbitrarily offset for visualization purposes. The step at $t \sim \SI{2}{s}$ is due to a flash of the OLIMPO calibration lamp. The spike at $t \sim \SI{7.6}{s}$ is due to a cosmic ray event, whose energy is distributed over the entire Si wafer of the array through ballistic and thermal phonons (see section \ref{cosmicraysevents}). Channels 9, 15, 17 are dark pixels.}
\phantomsection\label{fig:callamptimelines}
\end{figure}

\begin{figure}[htb]
\centering
\includegraphics[scale=0.72]{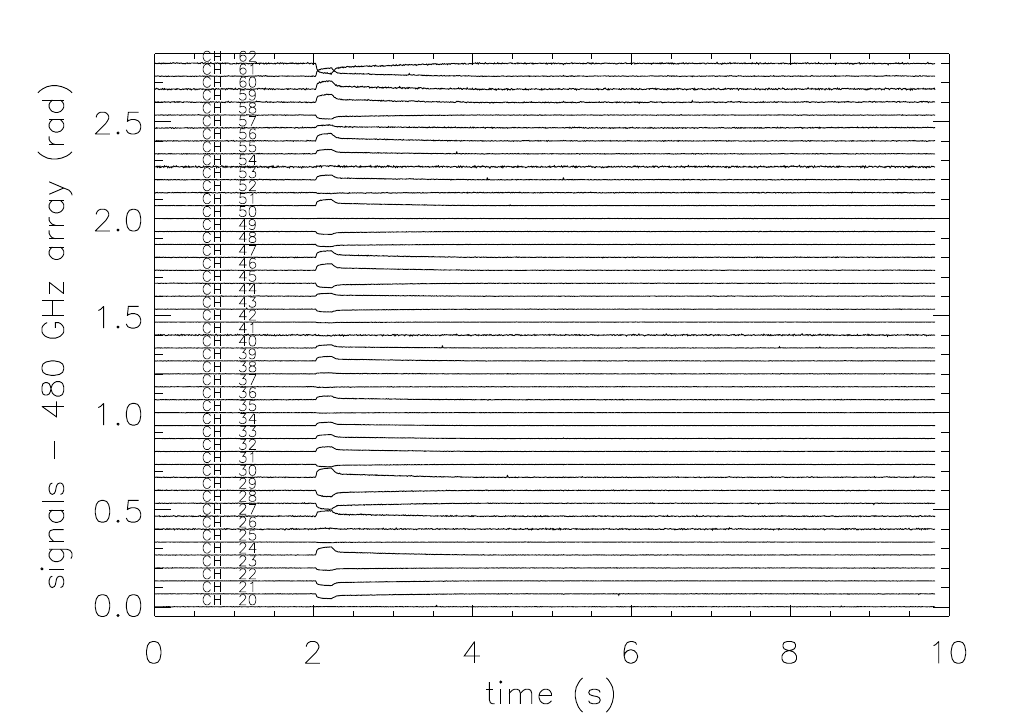}
\caption{\small A sample section of the raw signal tracks from 43 pixels of the \SI{460}{GHz} array, arbitrarily offset for visualization purposes. The step at $t \sim \SI{2}{s}$ is due to a flash of the OLIMPO calibration lamp. Channels 25 and 41 correspond to dark pixels.}
\phantomsection\label{fig:callamptimelines480}
\end{figure}

\begin{figure}[htb]
\centering
\includegraphics[scale=0.39]{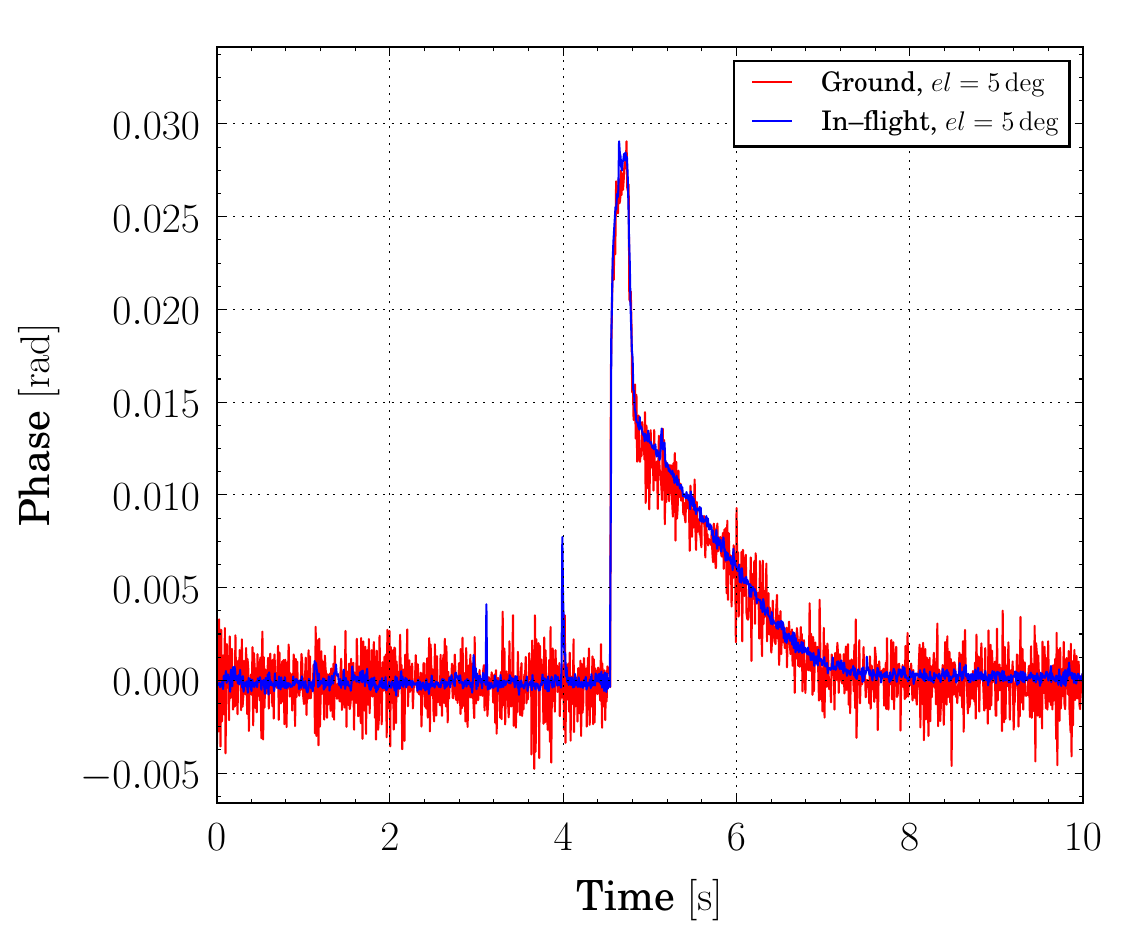}
\includegraphics[scale=0.39]{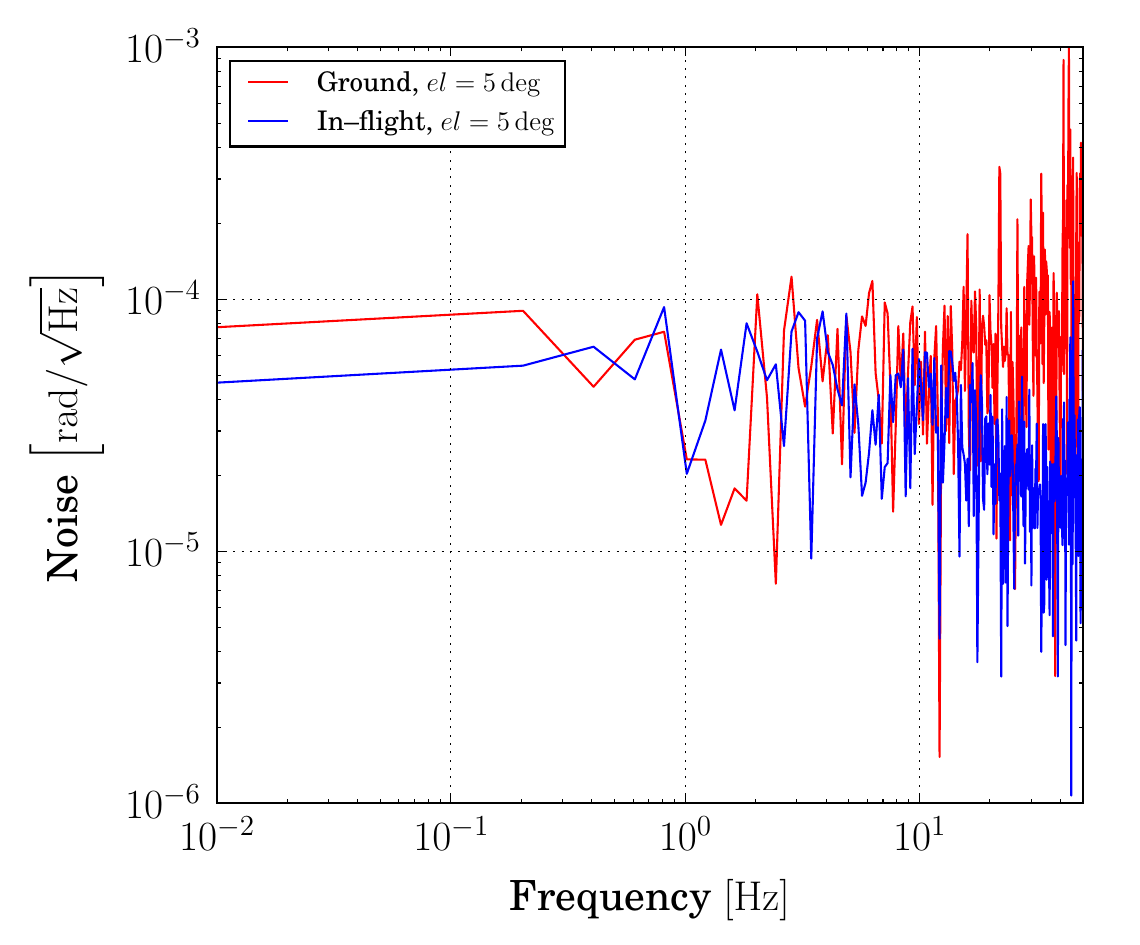}
\caption{\small \emph{Left panel}: comparison between the calibration lamp signals at ground (red line) and in--flight (blue line). The flight data have been renormalized to produce the same calibration lamp signal amplitude as the data taken on the ground. The improvement of the signal to noise ratio in flight is evident. \emph{Right panel}: comparison between the noise spectra at ground (red line) and in--flight (blue line). Both panels refer to pixel 7 of the \SI{150}{GHz} array.}
\phantomsection\label{fig:clamp_and_noise}
\end{figure}

In figures \ref{fig:callamptimelines} and \ref{fig:callamptimelines480} we report a short ($\sim$ \SI{10}{s}) representative section of the data from all pixels of the 150 and \SI{460} {GHz} arrays, which includes a calibration lamp flash at $t \sim \SI{2}{s}$. In the \SI{150}{GHz} array tracks, a large cosmic ray spike is evident ($t \sim \SI{7.6}{s}$). The shape of the calibration lamp signals (better visible in figure \ref{fig:clamp_and_noise}) with a sharp initial pulse followed by a slow linear decay, is due to the particular modulation profile selected for the current through the calibration lamp heater.


The response of different pixels to calibration lamp flashes depends on both the illumination from the calibration lamp and the selected working point in the resonance circle (which has been optimized as described in section \ref{tuning}). For most of the pixels the signal decreases when the photons flux increases, while increases for a few of them. The sign of the response to the energy deposition from cosmic rays is indeed in agreement with the sign of the response to the calibration lamp flash. Pixels 9, 15 and 17 in the \SI{150}{GHz} array and 25 and 41 in the \SI{460}{GHz} array are dark pixels and do not respond to calibration lamp flashes, while they respond to cosmic ray events. 

The performance of the detectors is not just a matter of responsivity: noise must also be taken into account. In order to estimate the detector performance improvement at float with respect to the ground, the in--flight detector signals produced by the calibration lamp have been normalized to the ones produced by the calibration lamp at ground. In this way, the in--flight and the ground responsivities become the same, but the noise levels are different. As an example, figure \ref{fig:clamp_and_noise} shows the comparison between the calibration lamp signals (\emph{left panel}), and the noise power spectra (\emph{right panel}), at ground and in--flight. This is for pixel \# 7 of the \SI{150}{GHz} array, but is fully representative of all pixels. 

As one can notice from the \emph{right panel}, the in--flight noise level is visibly reduced with respect to the noise at ground. So the in--flight NET is smaller than the NET on the ground, which means that the in--flight performance is better than on the ground. Quantitatively, the array--average reduction of the NET at \SI{12}{Hz} at float is a factor of 2.5 for the \SI{150}{GHz} array, 8 for the \SI{250}{GHz} array, 3.5 for the \SI{350}{GHz} array, and 5 for the \SI{460}{GHz} array (see table \ref{tab:NET_flight}). Several reasons contribute to these significant improvements: more accurate tuning, reduction of the radiative background, reduction of the level of microvibrations, reduction of the level of RFI (Radio Frequency Interference). As a result, the in-flight NET values are very close to the photon--noise--limited performance expected for the in--flight radiative environment of the OLIMPO photometric configuration.

\begin{table}[htb]
	\centering
		\fontsize{10pt}{18pt}\selectfont{
		\begin{tabular}{c|c|C|C}
		\hline
		\hline
		\multicolumn{1}{c|}{\multirow{1}{*}{Channel}}&
		\multicolumn{1}{c|}{\multirow{2}{*}{$N_{g}/N_{f}$}}&
		\multicolumn{2}{c}{\multirow{1}{*}{average NET$_{\rm RJ}$ $\left[\SI{}{\mu K\sqrt{\rm s}}\right]$}}\\
		\cline{3-4}
		\multicolumn{1}{c|}{\multirow{1}{*}{$\left[\SI{}{GHz}\right]$}}&
		\multicolumn{1}{c|}{\multirow{1}{*}{}}&
		\multicolumn{1}{c|}{\multirow{1}{*}{ground \citep{Paiella2019}}}&
		\multicolumn{1}{c}{\multirow{1}{*}{in--flight}}\\
		\hline
		\hline
150&2.5&201&81\\
250&8& 243&30\\
350&3.5& 243&69\\
460&5&336&67\\
\hline
		\hline
		\end{tabular}		
		}
		\caption{\small Array--averages of the ratio between the noise at ground ($N_{g}$) and in--flight ($N_{f}$), and noise equivalent temperature (in the Rayleigh--Jeans approximation, referred to the cryostat window, at a signal frequency of \SI{12}{Hz}) at ground and in--flight.}
	\phantomsection\label{tab:NET_flight}
\end{table}

\section{Cosmic ray events} \label{cosmicraysevents}

A concern for the operation of KIDs in space is the level of disturbance produced by cosmic ray (CR) hits on the detector array. The main interactions are ionization and atomic excitation due to charged particles (mainly protons, in primary cosmic rays) crossing the silicon wafer. The deposited energy produces ballistic phonons, spreading from the hit site towards the entire wafer. Ballistic phonons are fast, and thermalize below the Cooper pair binding energy in $\sim \SI{100}{\mu s}$, a timescale similar to the typical quasiparticle recombination time. For this reason the dead time due to CR events is usually smaller than in bolometric detectors, where the time constant is in the \SI{10}{ms} range (see e.g. \cite{Planck_CR, Maria_QUBIC}). In the OLIMPO KID arrays, a layer of superconductive Al has been added on the back side of the wafer, with the purpose of reducing the energy of ballistic phonons propagating in the wafer \cite{backshort1,backshort2} (in addition to acting as a backshort).   

\subsection{Non--thermal effects} \label{nontherm}

In silicon, the energy loss of protons with GeV energy is $dE/dx \sim \SI{0.23}{keV/\mu m}$, to first order irrespective of proton energy in the range of interest here (see e.g. \citep{2011JInst...6.6013M}). This deposited energy is converted in phonons, which propagate in the wafer and are detected by the resonators deposited on the wafer surface, producing fast spikes in the detector timestreams. The phase--variation signal produced by one of these events is
\begin{equation}
\Delta \phi = {\cal R}_\phi \eta  {dE \over dx} {x \over \tau_{qp} }
\label{eq:amplitude}
\end{equation}
where ${\cal R}_\phi \sim \SI{2e12}{rad/W}$ is the phase responsivity,  $Q = x \left(dE/dx\right) \sim \SI{3e-15}{J}$ is the energy loss in a $x \sim \SI{100}{\mu m}$ wafer thickness, $\tau_{qp}$ is the quasiparticle lifetime, and $\eta$ is the conversion efficiency from ionization energy into phonon energy and finally into energy for breaking Cooper pairs
(see e.g. \citep{Day, 2017ApPhL.110c3504C} and references therein).  Since the flight data samples consist of averages over $\Delta t \sim \SI{8}{ms}$ periods, with $\Delta t \gg \tau_{qp}$, the measured signals are diluted by a ratio $\sim \tau_{qp} / \Delta t$. So we expect a typical amplitude for the phase signal samples 
\begin{equation}
S = \Delta \phi {\tau_{qp} \over \Delta t }= {\cal R}_\phi \eta  {dE \over dx} {x \over {\Delta t} } \sim \eta \ {\rm rad}
\end{equation}
with $\eta \ll 1$. 

\subsection{Thermal effects} \label{therm}

In addition to the above described response, we have to consider that part of the energy from a CR crossing the Si wafer is dissipated thermally. If there is a non--negligible thermal resistance between the CR hit site and the thermostat (the $^3$He evaporator, in the case of OLIMPO), a thermal pulse will be induced in the temperature of the wafer and of the resonators. This will generate a frequency shift of the resonance, which will also be detected. Our design is aimed at minimizing this effect.  

The size of the thermal effect depends on the internal thermal resistance of the Si wafer, the contact resistance between the wafer and its metal holder, in parallel to the thermal resistance of the wire--bondings present between the wafer and its metal holder. 

At $T \sim \SI{0.3}{K}$, the internal conductivity of the Si wafer is  $k_{Si} \sim \SI{0.1}{W/m/K}$ (see {\it e.g.} \citep{1964PhRv..134.1058G}), while the specific heat capacity is $C_{Si} \sim \SI{6e-4}{J/K/kg}$ (see {\it e.g.} \citep{1959PhRv..113...33K}). From these values, using the heat diffusion equation, we can estimate the typical heat diffusion time over a distance $D$: 
\begin{equation}
\tau=\frac{D^{2}\rho_{Si}C_{Si}}{4k_{Si}}\;.
\end{equation}
In our case $D$ should be the distance between the CR hit site and the farthest pixel of the wafer, so we get $\tau \sim \SI{1.5}{ms}$. For a wafer thickness $x \sim \SI{0.1}{mm}$, the amplitude of the temperature pulse due to a typical energy release $Q \sim \SI{50}{keV}$ is very small: from the heat diffusion equation, where $\rho_{Si}$ is the mass density of the Si wafer and e is Euler's number:
\begin{equation}
\Delta T_{max}=\frac{Q/x}{\rho_{Si}C_{Si}D^{2}\pi{\rm e}}\sim\SI{20}{nK}\;.
\end{equation}

This temperature variation does not produce a significant frequency shift in the KID resonators: the phase signal produced by such a temperature change would be around $10^{-5} rad$, smaller than the noise.

The wafer is thermally connected to the holder at constant temperature by the contact over its back surface, and by tens of Al wire--bondings made with \SI{1}{mil} diameter wire, about \SI{3}{mm} long. The conductivity of 40 such wire--bondings is $\SI{10}{\mu W/K}< k_{b} < \SI{10}{mW/K}$ (strongly depending on the purity of the Al wire, see e.g. \cite{Wood}). The 3'' Si wafer has a heat capacity $V\rho_{Si}C_{Si} \sim \SI{8e-7}{J/K}$, so the time constant of the bolometric response of the entire wafer, with volume $V$, is $\tau \sim V\rho_{Si}C_{Si}/k_{b}$. Therefore $\SI{e-4}{s} < \tau < \SI{e-1}{s}$. For wires with average conductivity properties, this time constant is longer than the non--thermal one. In any case, the temperature fluctuation is $\Delta T_{max} = Q/\left(V\rho_{Si}C_{Si}\right) \sim \SI{10}{nK}$, so this kind of thermal effect is negligible in amplitude.

The contact thermal resistance is difficult to estimate, since it depends on the force pressing the wafer against the holder. In OLIMPO we use four teflon washers pressing the wafer (see \citep{Paiella2019}). The force can not be too strong, because we can not risk that the wafer breaks due to thermal contractions during the cooldown. Would the total conductivity be larger than $k_b$, given the heat capacity of the wafer estimated above, the resulting bolometric response would be faster than in the previous estimate, but would be the same in amplitude, again resulting in a negligible effect on the frequency shift of the KID resonators. 

The estimates above are quite uncertain, since depend on parameters which can not be fully controlled (contact conductivity, wire purity, heat capacity and internal conductivity of the wafer). The best approach is to measure the CR spikes and see if there are or not thermal effects. In fact, thermal effects tend to be slower than non--thermal ones. For the non--thermal effects, the recovery time is of the order of the recombination time for the quasiparticles, i.e. $\sim \SI{100}{\mu s}$ in our case \citep{Paiella2019}. This is shorter than our $\sim \SI{8}{ms}$ integration/sampling time of the flight readout electronics. The detection of decay times longer than the integration/sampling time would indicate the presence of thermal effects. 

\subsection{Events analysis}

\begin{figure}[htb]
\centering
\includegraphics[scale=0.86]{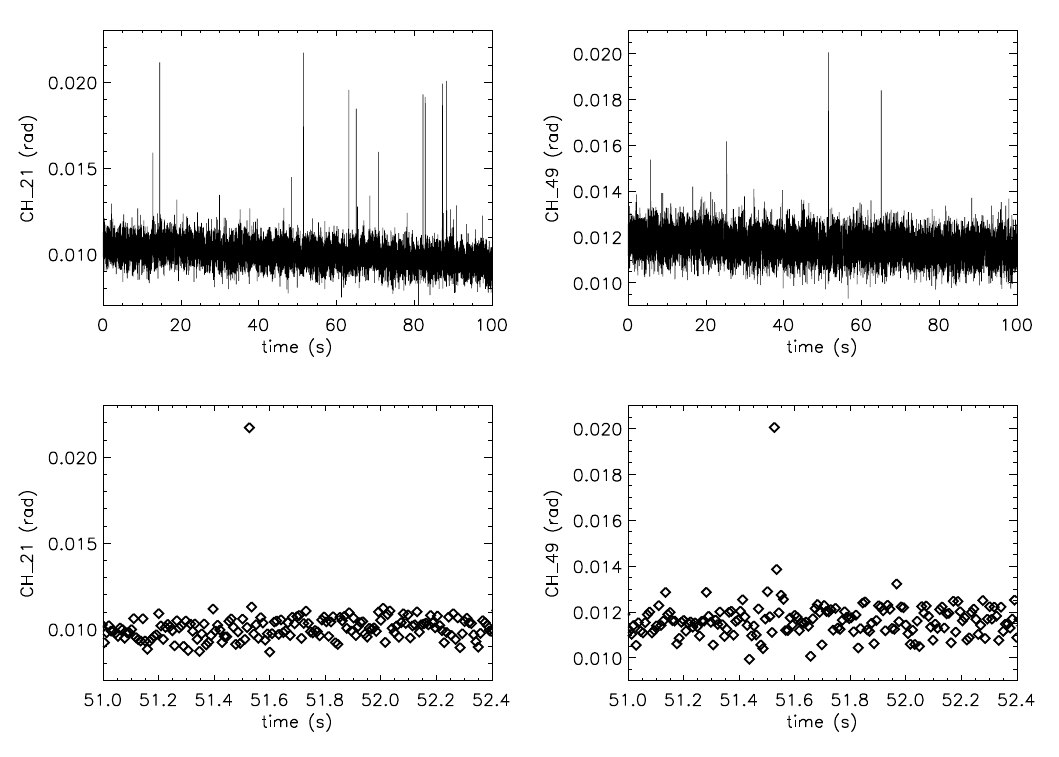}
\caption{\small \emph{Top row}: sample time--streams acquired at \SI{37.8}{km} of altitude from two KIDs of the \SI{460}{GHz} array, for the center pixel (channel 21, \emph{left panel}) and the next--to--center pixel (channel 49, \emph{right panel}).  \emph{Bottom row}: Zoom on a single spike in channels 21 (\emph{left}) and channel 49 (\emph{right}): the CR spikes are short (with a decay time $\tau$ shorter than the 8 ms integration time).}
\phantomsection\label{fig:spikestimelines480}
\end{figure}

\begin{figure}[htb]
\centering
\includegraphics[scale=0.86]{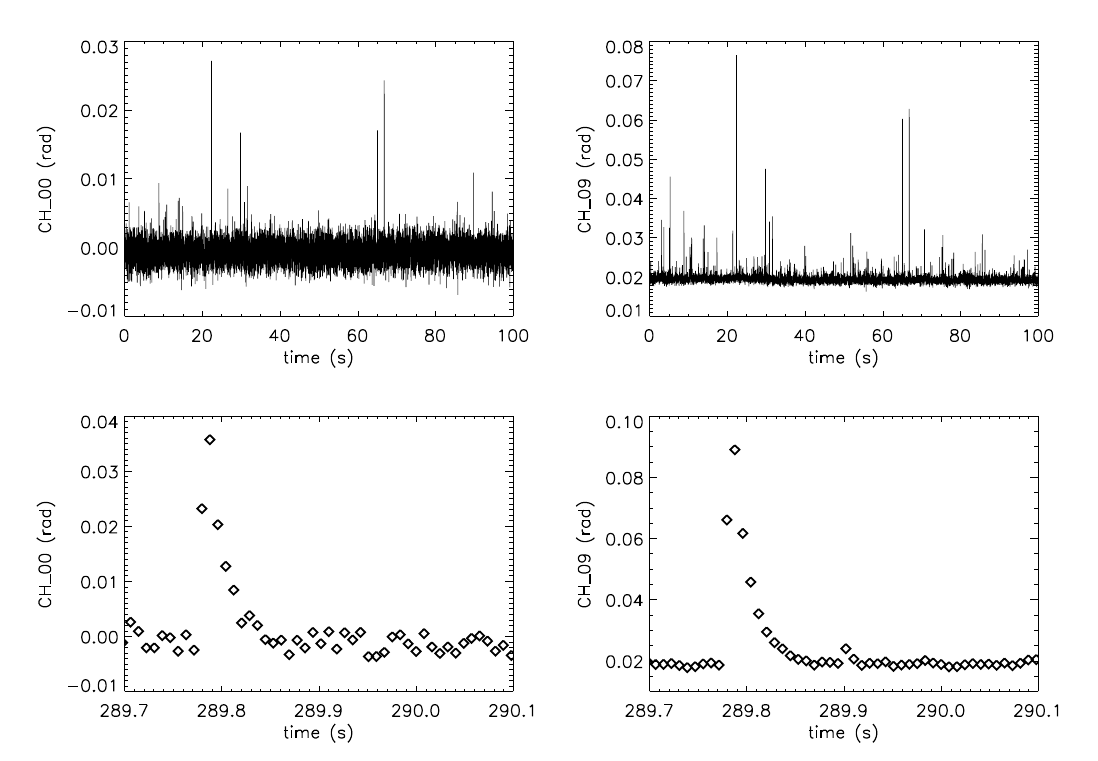}
\caption{\small \emph{Top row}: Sample signal streams acquired at \SI{37.8}{km} of altitude from two KIDs of the \SI{150}{GHz} array. One detector is exposed to mm--wave radiation from the sky (Channel 0, \emph{left panel}) and one is blanked (Channel 9, \emph{right panel}). The sampling frequency is \SI{122}{Hz}. CR induced spikes exceeding the noise level are evident as large positive deviations. Most of them are correlated between the two pixels.
\emph{Bottom row}: Zoom on a single spike in channels 0 (\emph{left}) and channel 9 (\emph{right}). The decay time ($\tau \sim 17$ ms) is determined by thermal effects, and is much longer than the detector quasiparticle lifetime.}
\phantomsection\label{fig:spikestimelines}
\end{figure}

In figure \ref{fig:spikestimelines480} we report sample timestreams for the center pixel and the next--to--center pixel of the \SI{460}{GHz} array. The period shown in figure \ref{fig:spikestimelines} is the first part of a $\sim$ \SI{820}{s} period (100000 samples) with stable detector bias and telescope pointing, so that sky signals are constant and noise is stationary. The full \SI{820}{s} period will be used for the analysis described below. 

A few large spikes are evident; smaller spikes are embedded in the noise. The order of magnitude of the spikes amplitudes is consistent with the estimate from equation \ref{eq:amplitude}. The fraction of data contaminated by evident cosmic rays hits is small (see below, sections \ref{large} and \ref{distribution}). Most of the hits are visible only on a single pixel, or in just two contiguous pixels, most likely because phonon propagation is damped by the superconducting layer on the back side of the wafer. The decay time is faster than the 8 ms integration time, pointing to the absence of thermal effects. The behaviour of the 250 and \SI{350}{GHz} arrays is very similar to the one of the \SI{460}{GHz} array, with no hint of thermal effects. 

In figure \ref{fig:spikestimelines} we display sample timestreams from a detector exposed to radiation (Channel 0) and one blanked (Channel 9), in the \SI{150}{GHz} wafer. For this array the CR hits rate is a bit higher than for the others (see sections \ref{large} and \ref{distribution}), and, at variance with the other arrays, the decay time is longer than the integration time; moreover, the spikes from a single CR hit are propagated to most of the pixels of the array. 

\begin{figure}[htb]
\centering
\includegraphics[scale=0.9]{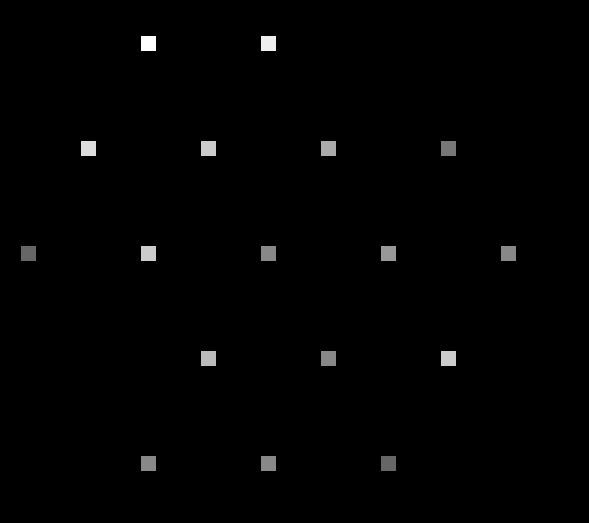}
\caption{\small Response of all the active pixels of the \SI{150}{GHz} channel to a single CR event (the same of the \emph{bottom row} of figure \ref{fig:spikestimelines}). The spike amplitude of each pixel is represented in grey levels. The CR particle has crossed the array in the \emph{top left} part, and the deposited energy has been distributed across the entire array, so that all the pixels detect the event.}
\phantomsection\label{fig:cosmicraymap}
\end{figure}

For example, the pixels labelled channel 0 and channel 9 are physically separated by $\sim$ \SI{35}{mm} on the wafer, and feature correlated spikes, as evident in figure \ref{fig:spikestimelines}. A sample map of the response of different pixels to the same cosmic ray is shown in figure \ref{fig:cosmicraymap}: it is evident that the deposited energy is spread over most of the wafer volume. Our sampling rate is not fast enough to detect delays in the spikes, depending on the relative position of the pixels and the CR hit. 

For all pixels of this array, each CR spike has a decay time $\tau \sim 17$ ms, slower than the quasiparticles recombination time, the integration time, and the sampling time. 

All this points to the presence of thermal effects, dominating the decay time of the pulses. Thermal effects of this kind had not been detected during calibrations in the lab, and have not been detected on extensive measurements carried out on the flight spare 150 GHz wafer after the flight. We can only suppose that the flight 150 GHz wafer had, for unknown reasons, a bad thermal contact with its holder, and a heat capacity smaller than our estimates. In any case, even if the response to cosmic rays is slower than expected, this fact does not affect significantly the performance of the array for the optical measurements.    

\subsection{Large cosmic ray spikes} \label{large}

\begin{figure}[htb]
\centering
\includegraphics[scale=0.45]{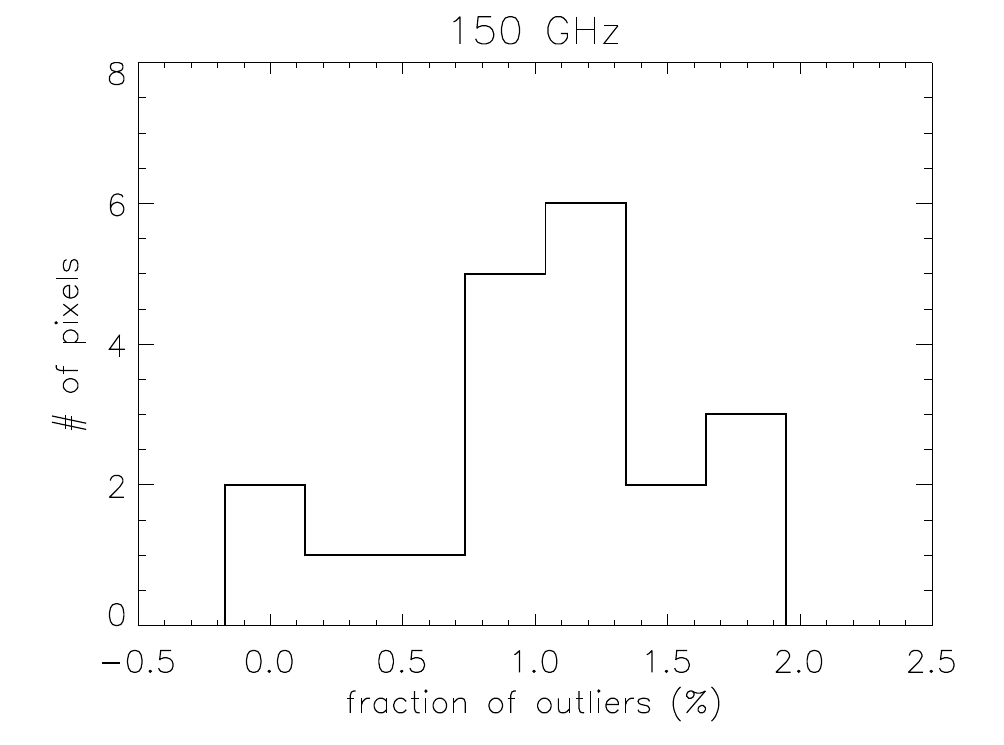}
\includegraphics[scale=0.45]{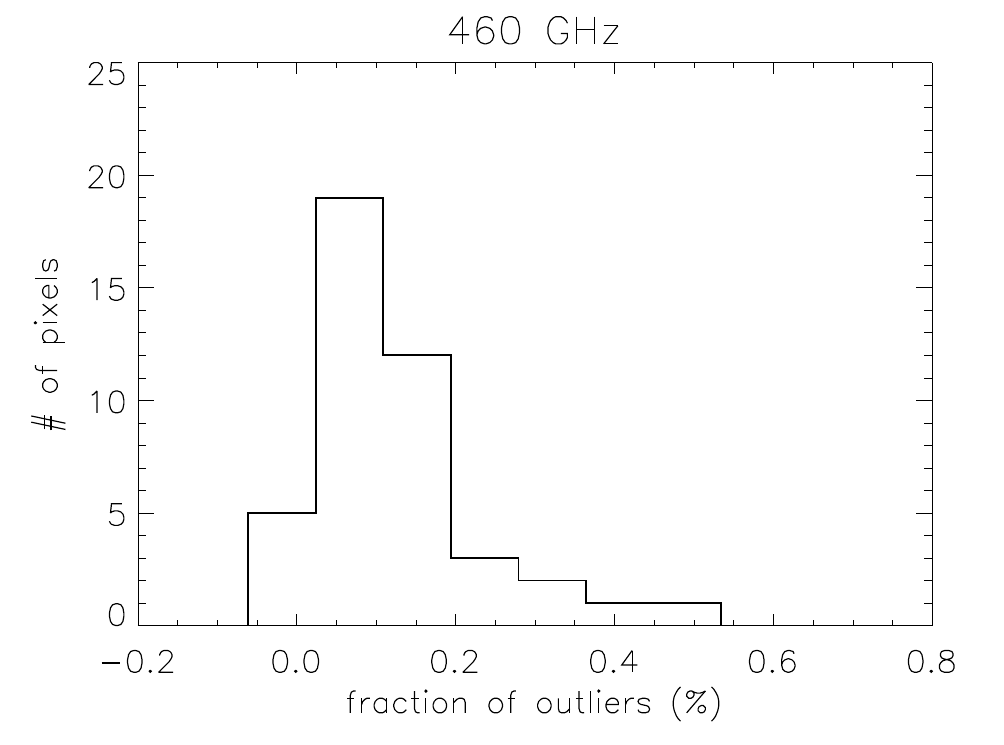}
\caption{\small Histograms of the number of array pixels featuring a given fraction of $> +3\sigma$ outliers (out of a total of 10$^5$ samples) for the \SI{150}{GHz} array (\emph{left}) and the \SI{460}{GHz} array (\emph{right}).}
\phantomsection\label{fig:spikesvspixel}
\end{figure}

Large spikes are easily identified in the timestreams. As a first estimate of the size of the disturbance in the data from large CR hits, we have estimated the standard deviation $\sigma$ of the time stream, in an iterative process where outliers at more than 3$\sigma$ are excluded from the estimate. Before each iteration step, we high--pass filter the data with a very long time constant (\SI{10}{s}) in order to remove the average value of the data, which is not of interest here. After convergence to $\sigma_c$ (the value of $\sigma$ obtained when on the next iteration $\sigma$ changes less than 1\%), we have simply considered the number of samples larger than the 3$\sigma_c$ level, and compared them to the total number of collected samples. In order to further correct for the expected presence of noise fluctuations at more than 3$\sigma$, unrelated to CR hits, we have evaluated the fraction of data smaller than $-3\sigma_c$ level, and subtracted this fraction from the fraction of data larger than $3\sigma_c$, for a rough noise de--bias of the estimate of the fraction of data contaminated by large CR hits. The results of this analysis are summarized in table \ref{tab:largespikes} and in figure \ref{fig:spikesvspixel}. 

\begin{table}[htb]
	\begin{center}
		\fontsize{10pt}{16pt}\selectfont{
		\begin{tabular}{l|r|r|r}
		\hline
		\hline
      \multirow{2}{*}{Array} & Working & \multirow{2}{*}{$f(>3\sigma_c)$} & \multirow{2}{*}{$f(>3\sigma_c)-f(<-3\sigma_c)$} \\
        & pixels  &  &  \\
		\hline
		\hline
		\SI{150}{GHz} & 20 & 1.5\% & 1.2\% \\
        \SI{250}{GHz} & 34 & 1.0\% & 0.4\% \\
        \SI{350}{GHz} & 23 & 0.4\% & 0.1\% \\
		\SI{460}{GHz} & 43 & 0.4\% & 0.2\% \\
		\hline
		\hline
		\end{tabular}
		}
		\caption{\small Fraction of samples with values $>3\sigma_c$ in a \SI{820}{s} long stable--pointing period (total 100000 samples per pixel), for the four arrays of the experiment. To compute the fractions, we coadded all the events of all the pixels of the same array. The fraction of data contaminated by CR hits is small for all the pixels of all the arrays.}
		\phantomsection\label{tab:largespikes}
\end{center}
\end{table}

The fraction of data contaminated by large CR spikes is small for all active pixels in all the OLIMPO arrays. The fraction of contaminated data is larger for the \SI{150}{GHz} array than for the other arrays, since in the \SI{150}{GHz} array the energy from CR hits is spread over most of the pixels, while in the other arrays the energy from each hit is propagated only to the pixels closest to the hit site. The \SI{250}{GHz} channel is known to be affected by low--level microphonic noise, so, for this array, the quoted fractions should be considered upper limits.

\subsection{Distributions of cosmic ray spikes} \label{distribution}

While the analysis above is assessing the issue of CR hits contamination in the OLIMPO data, is not trivial to generalize, because it is strictly related to the noise level of the OLIMPO detectors: we have just estimated how many samples are contaminated at a level larger than the noise of the OLIMPO detectors.

In order to get more general insight on the level of primary CR hits in space, and also to estimate the contamination due to CR embedded in the noise but still affecting the sensitivity of the system (see e.g. \citep{2010AA...519A..24M}), we developed a simple model of the interaction of CR with silicon wafers, and fitted our data to the model. In practice we histogram the data, and fit the histogram with the superposition of Gaussian noise plus a simple model for the distribution of CR spikes. 

To derive the analytical form of the spike amplitude distribution, we assume that the spike amplitude $S$ is proportional to the deposited energy, which in turn is proportional to the length of the ionizing particle path crossing the silicon wafer. This is a solid disk, with a thickness ranging between 100 and $\SI{300}{\mu m}$ and a diameter of 2 or 3 inches for the OLIMPO arrays. It is thus reasonable to model the wafer as an infinite slab, neglecting edge effects. Under these hypotheses
\begin{equation}
S = \frac{k x}{\cos \theta}\;,
\end{equation}
where $\theta$ is the incidence angle of the incoming particles with respect to the normal to the wafer surface, $x$ the thickness of the silicon wafer, and $k$ is a proportionality constant: 
\begin{equation}
k = \frac{{\cal R}_\phi \eta}{ \Delta t} \frac{dE}{dx}\;.
\end{equation}
So we have:
\begin{equation}
\begin{split}
\frac{dN}{dS} &= \frac{dN}{dA d\theta} \left|\frac{d\theta}{dS}\right| A_o \cos \theta = 2 \pi \sin \theta \frac{dN}{dA d\Omega} \left|\frac{d\theta}{dS}\right| A_o  \cos \theta =\\
&=2 \pi A_o \frac{dN}{dA d\Omega} \frac{\cos^3\theta}{k x } =2 \pi  A_o \frac{dN}{dA d\Omega} \frac{\left(k x\right)^2} {S^3}\;, 
\end{split}
\end{equation}
where $A_o$ is the area of the Si wafer where the pixels are deposited. In the limit where the incoming particles (primary cosmic rays) have isotropic velocities, this is a simple $1/S^3$ distribution, with a low--$S$ cut--off due to the minimum energy deposition at $S_{min}=kx$ and a high--$S$ cut--off due to the maximum possible energy deposition at $S_{max}=kD$, where $D$ is the diameter of the wafer. 

\begin{figure}[htb]
\centering
\includegraphics[scale=0.45]{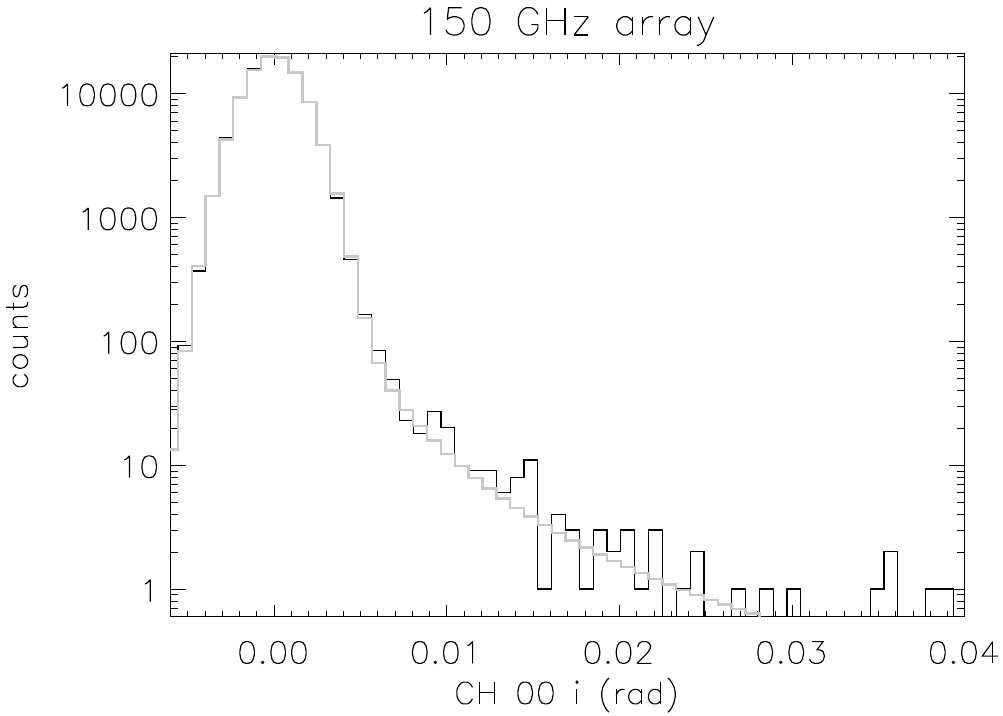}
\includegraphics[scale=0.45]{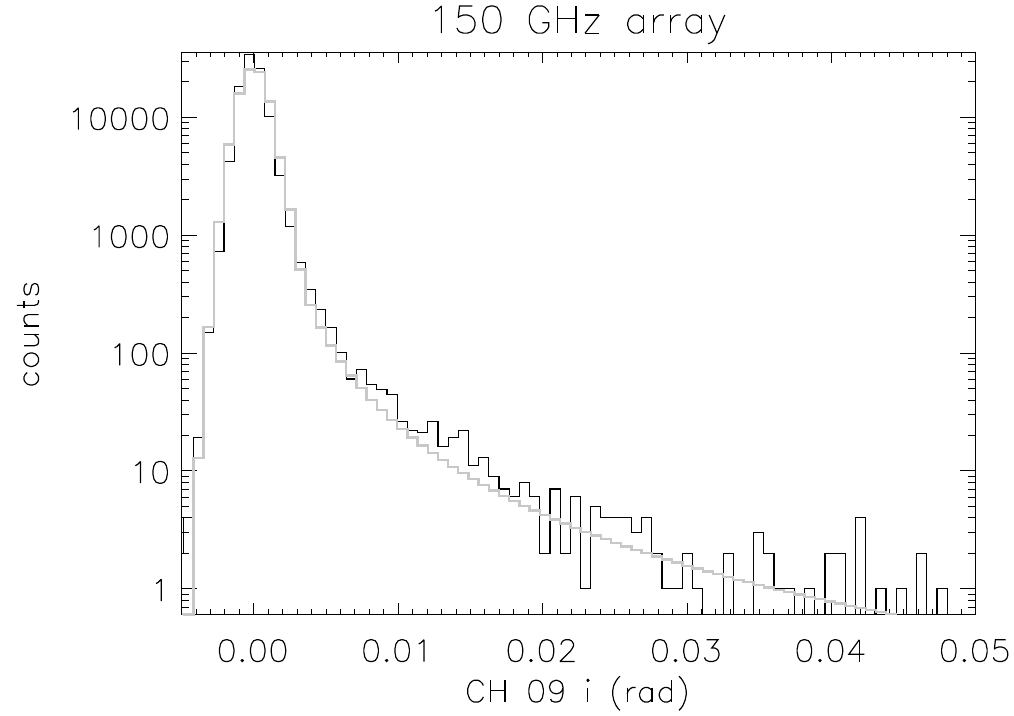}
\caption{\small Histograms of the sample \SI{820}{s} long in-phase (i) timestreams representative of the operation of two OLIMPO KIDs of the \SI{150}{GHz} array, in the absence of sky scans and with stable biasing (black line). The positive tail is due to CR hits. The Gaussian noise in the active channel (CH 00, \emph{left}) is larger than the Gaussian noise in the blank channel (CH 09, \emph{right}) due to photon noise. The lighter lines represent the best fits using the CR spike model described in the text and a Gaussian model for detector noise.}
\phantomsection\label{fig:spikehistograms150}
\end{figure}

\begin{figure}[htb]
\centering
\includegraphics[scale=0.45]{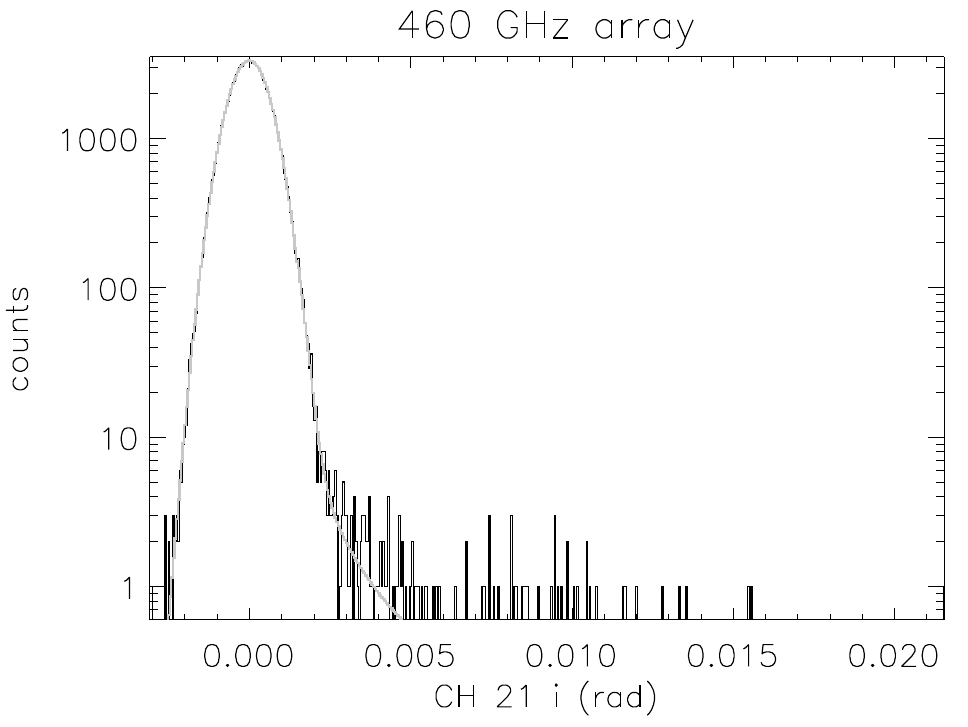}
\includegraphics[scale=0.45]{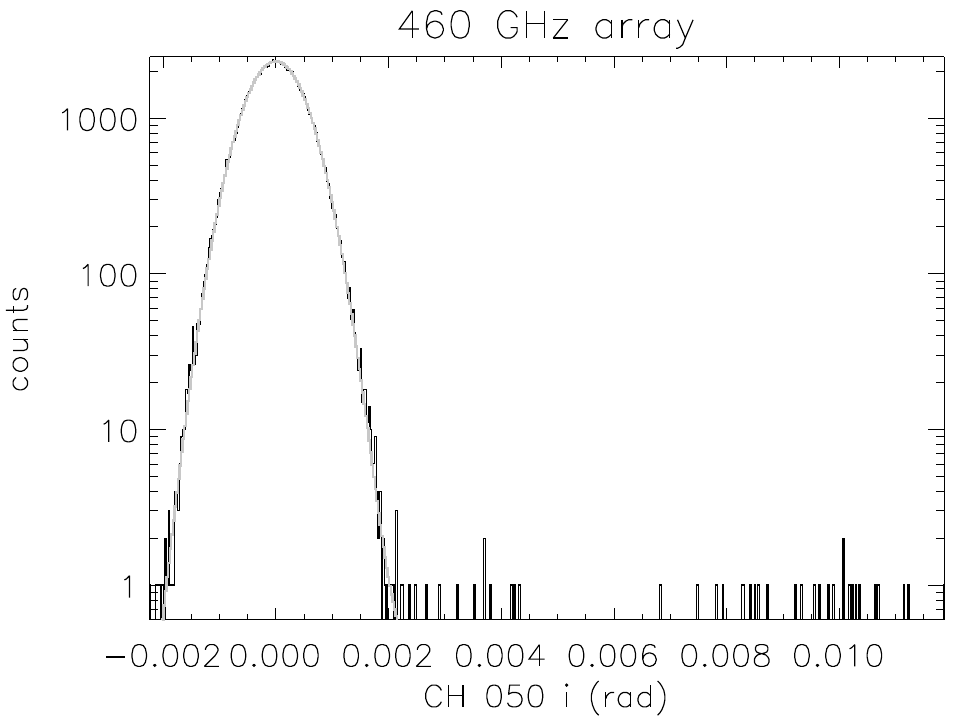}
\caption{\small Histograms of the sample \SI{820}{s} long in-phase (i) timestreams representative of the operation of two OLIMPO KIDs of the \SI{460}{GHz} array, in the absence of sky scans and with stable biasing (black line). The Gaussian noise in the active channel (CH 21, \emph{left}) is larger than the Gaussian noise in the blank channel (CH 50, \emph{right}) due to photon noise. The lighter lines represent the best fits using the CR spike model described in the text and a Gaussian model for detector noise. }
\phantomsection\label{fig:spikehistograms460}
\end{figure}

In figure \ref{fig:spikehistograms150} we report the histograms of sample KID signals for the \SI{150}{GHz} array, for a period of \SI{820}{s}, after the average removal, and the best fits for the histograms using Gaussian noise plus the distribution of spikes described above. In figure \ref{fig:spikehistograms460} we report the same for the \SI{460}{GHz} array. The number of spikes per pixel islower than in the \SI{150}{GHz} case, due to limited energy diffusion around the CR hit site, so that only one or just a few pixels are affected by the event. The 250 and \SI{350}{GHz} arrays perform basically like the \SI{460}{GHz} array.

\begin{figure}[htb]
\centering
\includegraphics[scale=0.43]{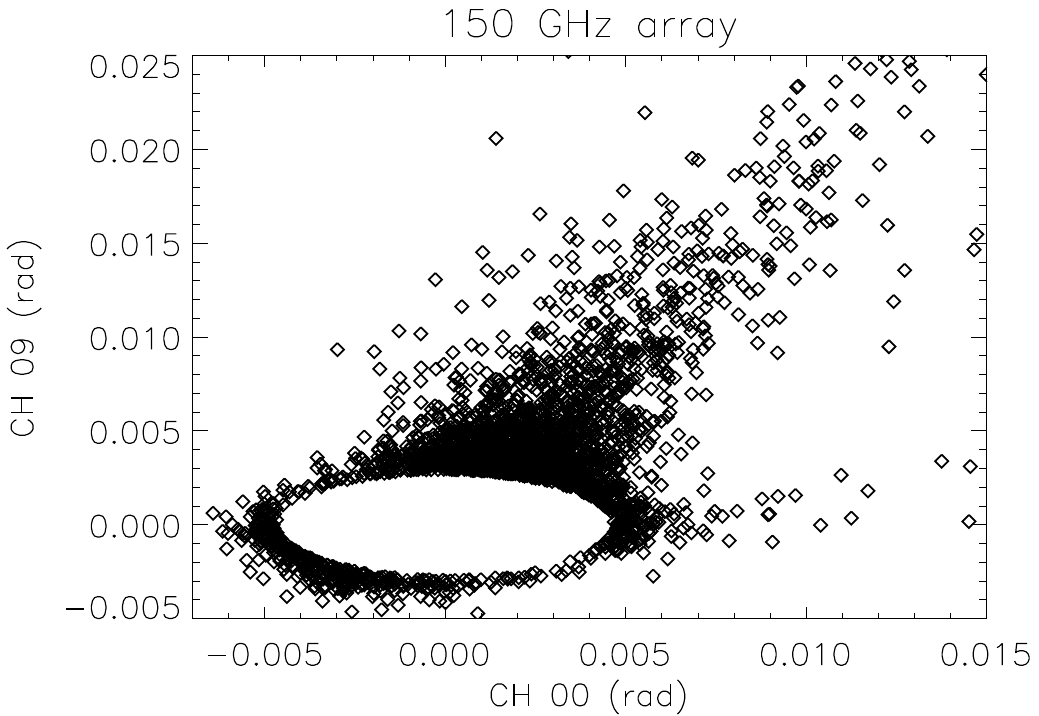}
\includegraphics[scale=0.43]{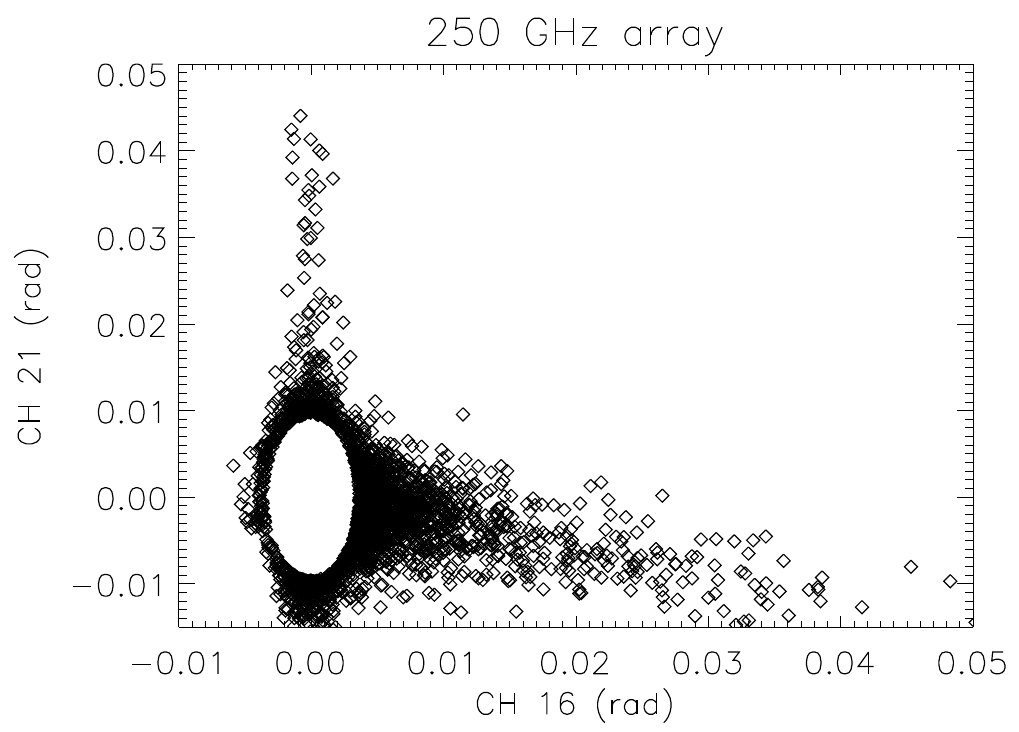}
\includegraphics[scale=0.43]{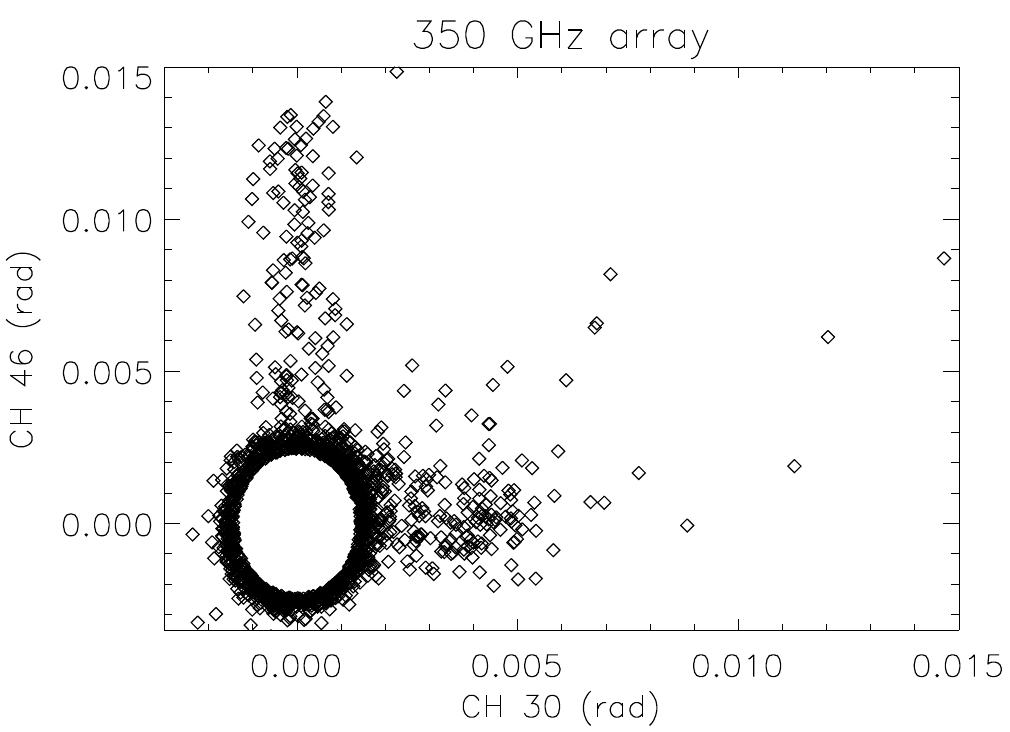}
\includegraphics[scale=0.43]{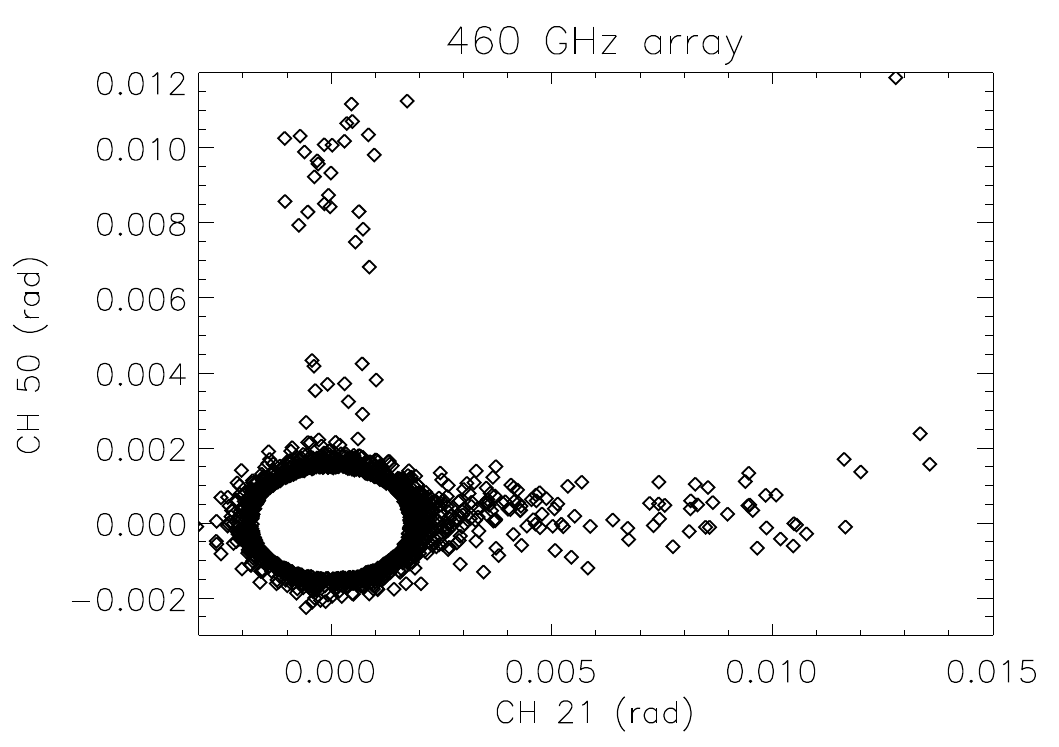}

\caption{\small Scatter plots of the signals from the dark pixel versus the signal from the center pixel, for all the arrays of the OLIMPO experiment. Data points distant less than $r=2\sqrt{\sigma_x^2+\sigma_y^2}$ from the origin have not been plotted.}
\phantomsection\label{fig:scatterplots}
\end{figure}

This different behaviour is evident from the scatter plots of the signals from the center and dark pixels (see figure \ref{fig:scatterplots}), which in the \SI{150}{GHz} array display a clear correlation of the spikes, even if the considered pixels (CH 00 and CH 09) are separated by a physical distance of \SI{35}{mm}. In the other arrays such a correlation is present only for neighbouring pixels while is absent for distant pixels.

We use the fits to the histograms to estimate the total fraction of data contaminated by CR events. We integrate the best fit of the spikes amplitude distributions in the range between the best fit minimum amplitude $S_{min}$ and the maximum detected spike amplitude. 

The results are reported in table \ref{tab:spikes} below, and confirm that the contamination from CR hits is relatively modest for all the pixels. 

\begin{table}[htb]
	\begin{center}
		\fontsize{10pt}{16pt}\selectfont{
		\begin{tabular}{l|r|r}
		\hline
		\hline
        \multirow{2}{*}{Pixel / Array} &  \multirow{2}{*}{$S_{min}$ $\left[rad\right]$} &fraction of data \\
        & & contaminated by CR  \\
		\hline
		\hline
		00 (active) / \SI{150}{GHz} & $(3.6 \pm 0.3) \times 10^{-3}$ & $[0.53\substack{+0.22 \\ -0.19}] \%$ \\
		09 (dark) / \SI{150}{GHz}  & $(2.5 \pm 0.1) \times 10^{-3}$ & $[2.73\substack{0.14 \\ -1.10}] \%$ \\
		16 (active) / \SI{250}{GHz} & $(2.4 \pm 0.1) \times 10^{-3}$ & $[2.84\substack{+0.49 \\ -0.49}] \%$  \\
		21 (dark) / \SI{250}{GHz}  & $(13.7 \pm 0.8) \times 10^{-3}$ & $[0.075\substack{+0.019 \\ -0.038}] \%$  \\
		30 (active) / \SI{350}{GHz} & $(2.5 \pm 1.1) \times 10^{-3}$ &  $[0.123\substack{+0.085 \\ -0.059}] \%$ \\
		46 (dark) / \SI{350}{GHz}  & $(6.2 \pm 1.5) \times 10^{-3}$ & $[0.033\substack{+0.027 \\ -0.030}] \%$  \\
		21 (active) / \SI{460}{GHz} & $(1.7 \pm 0.9) \times 10^{-3}$ & $[0.16 \substack{+0.21 \\ -0.09}] \%$  \\
		50 (dark) / \SI{460}{GHz}  & $(1.3 \pm 5.6) \times 10^{-3}$ & $ < 0.013 \%$ \\  
		\hline
		\hline
		\end{tabular}
		}
		\caption{\small Best fit minimum signal $S_{min}$ and fraction of data contaminated by CR spikes, computed from the best fit to the theoretical distribution of the spikes, for the center active KID and one blanked KID, in the four OLIMPO KID arrays. The 68\% C.L. intervals specified take into account the correlations between the different fitted parameters.}
		\phantomsection\label{tab:spikes}
\end{center}
\end{table}

These numbers depend only on the wafer composition and geometry, and CR environment. We can assume that the CR background experienced by OLIMPO at 37.8 km of altitude is roughly representative of the primary cosmic rays background in low-Earth orbit (LEO), since most of the secondary interactions happen at lower altitudes. So, in a LEO satellite mission avoiding the radiation belts, with an instrument using similar arrays of KIDs, the effects of CRs hits will be similar to the ones measured here. Moreover, given the lower radiative background conditions of a satellite instrument, resulting in a lower photon-noise level for the detectors, it will be easier to remove CR spikes already in the time domain, since the spikes will clearly stand out of the reduced noise fluctuations. In the end, the fraction of contaminated data should be modest, as in table \ref{tab:spikes}. We can conclude that these results pave the way to the use of mm--wave KIDs in satellite missions. 

\section{Conclusions}

We have demonstrated experimentally that Kinetic Inductance Detectors can be used efficiently at stratospheric balloon altitude. In particular we have shown that the detectors readout can be tuned in flight, obtaining superior (photon-noise limited) performance with respect to the laboratory one; in addition we have shown that only very large background variations (as the insertion of a room--temperature spectrometer in the optical path) require re--tuning of the detector readout. Moreover, we have shown that cosmic rays hits do affect the noise performance of the detectors, but the contamination is limited to less than 4\% of the data for all pixels, and below 1\% of the data for most of the pixels. These results have been obtained with photon noise limited detectors operated in the mm windows at 150, 250, 350 and \SI{460}{GHz}, in the radiative background of the residual atmosphere, with a room--temperature telescope. Since the cosmic rays background at that altitude is dominated by primary cosmic rays, these results are also roughly representative of what one can expect in a LEO satellite mission, with similar KIDs and radiative environment. 

\acknowledgments{
\addcontentsline{toc}{section}{Acknowledgements}
We acknowledge the Italian Space Agency (ASI) for funding the OLIMPO payload and flight, and the development of Kinetic Inductance Detectors for space. We acknowledge Sapienza University of Rome for co--funding the OLIMPO experiment. We are grateful to Angelo Cruciani for useful discussions and suggestion, and to Giorgio Amico for precision machining of many mechanical parts. We warmly thank Mikko Syrj\"asuo of UNIS Longyearbyen and Anthony Moule of ELTA (ECA Group, Toulouse) for support with the line--of--sight telemetry.}

\newpage{}
\bibliography{inflight}
\bibliographystyle{JHEP} 
\addcontentsline{toc}{section}{References}

\end{document}